\documentclass[10pt,amsmath,amssymb,twocolumn,superscriptaddress,nofootinbib,aps,prl,twocolumn,floatfix,showkeys]{revtex4-2}

\usepackage{graphicx}  
\graphicspath{{./plots/}}
\usepackage{bm}        
\usepackage{enumitem}
\usepackage{etoolbox}
\usepackage{MnSymbol}

\usepackage[colorlinks=true]{hyperref}

\providecommand{\texorpdfstring}[2]{#1}

\newcommand{\comment}[1]{}

\newcommand{\lr}[1]{ \left( #1 \right) }
\newcommand{\lrs}[1]{ \left[ #1 \right] }
\newcommand{\lrc}[1]{ \left\{ #1 \right\} }
\newcommand{\vev}[1]{ \left\langle \, #1 \, \right\rangle }
\newcommand{\cev}[1]{ \left\llangle \, #1 \, \right\rrangle }

\newcommand{\MeV}[0]{ \, \mathrm{MeV}}

\newcommand{\tr}{ {\rm Tr} \, }
\newcommand{\im}{ {\rm Im} \, }
\newcommand{\re}{ {\rm Re} \, }

\newcommand{\ket}[1]{ \, | #1 \rangle }
\newcommand{\bra}[1]{ \langle #1 | \, }
\newcommand{\braket}[2]{ \langle #1 | #2 \rangle \, }

\newcommand{\expa}[1]{ \exp{\left( #1 \right)} }
\newcommand{\abs}[1]{| #1 |}

\newcommand{\OO}{\hat{\mathcal{O}}}
\newcommand{\HH}{\hat{\mathcal{H}}}
\newcommand{\EE}{\hat{\mathcal{E}}}

\newcommand{\hpsi}{\hat{\psi}}
\newcommand{\hpsid}{\hat{\psi}^{\dag}}
\newcommand{\htheta}{\hat{\theta}}
\newcommand{\hdd}{\hat{\Delta}}
\newcommand{\ZZ}{\mathcal{Z}}
\newcommand{\dtau}{{\Delta \tau}}
\newcommand{\dt}{{\Delta t}}

\begin{document}
\sloppy
	
\title{Spectral reconstruction based on dimensional reduction in high-temperature gauge theories}

\author{P. V. Buividovich}
\email{pavel.buividovich@liverpool.ac.uk}
\affiliation{Department of Mathematical Sciences, University of Liverpool, UK}

\author{B. Hind}
\email{ben.hind@liverpool.ac.uk}
\affiliation{Department of Mathematical Sciences, University of Liverpool, UK}

\date{December 29th, 2025}
\begin{abstract}
We propose a numerical spectral reconstruction workflow for high-temperature gauge theories that incorporates elements of semi-classical real-time evolution directly into standard lattice QCD simulations via high-temperature dimensional reduction, thus counteracting the deterioration of Euclidean-time correlators at high temperatures. With a moderate numerical cost, our method allows to estimate spectral functions with parametrically better frequency resolution as compared with spectral reconstruction methods based on Euclidean-time correlators alone. The method is tested on a simple $\lr{1+1}$-dimensional Abelian gauge theory with fermions, where our method precisely reproduces the full quantum spectral functions calculated using exact numerical diagonalization in the high-temperature, weak-coupling regime. We also demonstrate the feasibility of our approach by applying it to meson correlators in lattice QCD deep in the deconfinement regime.
\end{abstract}

\maketitle


\emph{Introduction.} Quantum Monte-Carlo (QMC) method is an important tool for first-principle studies of gauge theories, including Quantum ChromoDynamics (QCD). Most QMC simulations operate in Euclidean (imaginary) time and produce systematically improvable results for thermodynamic quantities such as the equation of state. However, extracting real-time responses such as electric conductivity or shear viscosity from QMC simulations is a mathematically ill-defined problem. It is often formulated as the inversion of the Green-Kubo relations
\begin{eqnarray}
\label{eq:GK0}
	G_E\lr{\tau} = \int\limits_0^{+\infty} d w \, \frac{\cosh\lr{w \lr{\tau - \beta/2}} }{\sinh\lr{w \beta/2}} \, S\lr{w}	,
\end{eqnarray}
where the input is the Euclidean-time correlator $G_E\lr{\tau} = \ZZ^{-1} \tr\lr{\OO e^{-\tau \HH} \OO e^{-\lr{\beta - \tau} \HH}}$ for some operator $\OO$, and the output is some approximation to the real-time spectral function $S\lr{w} = \ZZ^{-1} \sum\limits_{m,n} \abs{\bra{m} \OO \ket{n}}^2 \lr{e^{-\beta E_m} - e^{-\beta E_n}} \, \delta\lr{w - \lr{E_n - E_m}}$. Here $\HH$ is the Hamiltonian with eigensystem $\lrc{E_m, \ket{m}}$, $\ZZ = \tr e^{-\beta \HH}$, $\beta \equiv T^{-1}$ is the inverse temperature, and $\tau \in \lrs{0, \beta}$ is the Euclidean time.

Using the relation $S\lr{w} = \pi^{-1} \im \int dt\, e^{i w t} G_R\lr{t}$ between $S\lr{w}$ and the real-time retarded correlator $G_R\lr{t}$, one concludes that the sensitivity of Euclidean-time correlator $G_E\lr{\tau}$ to dynamics at real (Minkowski) time $t$ decays exponentially as $e^{-2 \pi T \, t}$. This explains the limited frequency resolution $\Delta w \gtrsim \pi \, T$ of the numerical estimates of $S\lr{w}$ obtained from spectral reconstruction methods which use $G_E\lr{\tau}$ as the only input, such as the Backus-Gilbert or the Maximal Entropy (MaxEnt) methods \cite{Gubernatis:96:1,Asakawa:hep-lat/0011040,Meyer:1104.3708,Tripolt:1801.10348,Ulybyshev:1707.04212,Ostmeyer:2510.15500,Lawrence:2408.11766,Bennett:2405.01388,Wagman:2406.20009}, as well as NP-hardness of achieving a parametrically better resolution without changing the workflow \cite{Rothkopf:1610.09531}. While functional methods provide an interesting alternative approach to this problem \cite{Smekal:1807.04952,Smekal:2112.12568,Schlichting:2009.04194,Pawlowski:2107.13464,Pawlowski:2210.07597,Fischer:1705.03207}, here we focus on QMC simulations on the lattice.

The scaling $\Delta w \sim \pi \, T$ of the frequency resolution makes numerical reconstruction of $S\lr{w}$ from $G_E\lr{\tau}$ increasingly hard at high temperatures. At the same time, asymptotically free gauge theories such as QCD become weakly coupled in this regime, which warrants the use of (semi-)classical techniques for spectral function calculations \cite{Aarts:hep-ph/0108125,Berges:0912.3135,Boguslavski:1804.01966,Smekal:2112.12568,Boguslavski:2106.11319}. Most of such calculations, however, focus on phenomenological initial states or non-thermal fixed points of real-time dynamics rather than on the linear-response description of small perturbations of thermal equilibrium state represented by first-principle Euclidean-time QMC simulations. 

In this paper we propose a spectral reconstruction workflow for high-$T$ gauge theories that combines the advantages of both QMC and semi-classical methods, and can be directly integrated into conventional lattice QCD simulations. We first describe our approach for a generic gauge theory, then compare its output with numerical exact diagonalization results for $\lr{1+1}$-dimensional $U\lr{1}$ gauge theory with fermions, and finally apply it to meson correlators in high-temperature lattice QCD. Technical details of our calculations are summarized in the Supplementary Material.


\emph{Dimensional reduction at high $T$ and spectral reconstruction.} Within the Euclidean path integral formulation, the emergence of the semi-classical high-temperature regime is equivalent to the suppression of all non-zero Matsubara modes, so that the gauge field $\vec{A}\lr{\tau, \vec{x}}$ becomes effectively independent of the Euclidean time $\tau$ \cite{GINSPARG1980388,AppelquistPisarskiPRD1981MagneticQCD,Braaten:hep-ph/9510408,Hands:hep-lat/0703001}. An equivalent statement is that the density matrix becomes effectively diagonal \cite{Greiner:hep-th/9605048,Stephanov:hep-ph/9502302,deCarvalho:hep-th/0101142} in the eigenbasis of $\vec{A}$ and can be represented as
\begin{eqnarray}
\label{eq:density_matrix}
	\hat{\rho}^s 
	=
	\int \mathcal{D}\vec{A} \, w\lrs{\vec{A}} \, \int \mathcal{D}\Omega \, 
	\hat{\Omega} \, \ket{\vec{A}} \bra{\vec{A}} \otimes e^{-\beta \hpsid h\lrs{\vec{A}} \hpsi} ,
\end{eqnarray}
where the integration is over all gauge fields $\vec{A}\lr{\vec{x}} = T \, \int_0^{\beta} d\tau \, \vec{A}\lr{\tau, \vec{x}}$ with the weight $w\lrs{\vec{A}}$ obtained by integrating out higher Matsubara modes of $\vec{A}\lr{\tau, \vec{x}}$ in the path integral. We refer to the averaging of $\vec{A}\lr{\tau, \vec{x}}$ over $\tau$ as ``static projection'' and use the superscript ``s'' for results of calculations with static gauge fields. The term $e^{-\beta \hpsid h\lrs{\vec{A}} \hpsi}$ is the density matrix for the fermion fields $\hpsi$, $\hpsid$ that couple to the classical gauge field $\vec{A}$ via their single-particle Hamiltonian $h\lrs{\vec{A}}$. $\hpsid h\lrs{\vec{A}} \hpsi$ is a short-hand notation for the bilinear form $\int d^d\vec{x} \, \hpsid_a\lr{\vec{x}} h_{ab}\lrs{\vec{A}\lr{\vec{x}}} \hpsi_b\lr{\vec{x}}$, where summation over $a$, $b$ accounts for internal fermion degrees of freedom (e.g. spin, colour, flavour). Integration over all gauge transformations $\Omega\lr{\vec{x}}$ acting on both fermion and gauge fields projects the density matrix to the physical Hilbert space. 

Consider now the Euclidean-time correlator of some fermionic bilinear operator $\OO = \hpsid O \hpsi$. Using the well-known representation of the fermionic Green's functions in terms of (time-ordered) exponentials of $h\lrs{\vec{A}}$ \cite{Blankenbecler:PhysRevD.24.2278} and assuming that the path integral is dominated by $\tau$-independent gauge fields $\vec{A}\lr{\vec{x}}$, we obtain for the contribution of fermionic connected diagrams
\begin{eqnarray}
\label{eq:GE_static}
	G_E^s\lr{\tau} 
	=
	\vev{
	\tr\lr{
		O \, n_z\lr{h} \, O\lr{\tau} \, n_z\lr{h} e^{-\beta h} z} + \bar{O}^2} , 
\end{eqnarray} 
where the trace is over the fermionic single-particle Hilbert space, $O\lr{\tau} = e^{-\tau \, h} \, O \, e^{\tau \, h}$, $\vev{\ldots}$ denotes averaging over $\vec{A}$ and $z$, $\bar{O} = \tr\lr{O \, n_z\lr{h}}$ is a $\tau$-independent disconnected contribution, and $n_z\lr{h} = \lr{1 + e^{-\beta h} z}$ is the center-modified Fermi distribution function. While the distribution of $\vec{A}$ is the same as in (\ref{eq:density_matrix}), the factor $\bra{\vec{A}} \hat{\Omega} \ket{\vec{A}}$ in the trace over the bosonic Hilbert space restricts $\Omega$ to gauge transformations which leave any vector potential $\vec{A}\lr{\vec{x}}$ unchanged. Such transformations are generated by coordinate-independent elements of the group center $\Omega\lr{\vec{x}} = z$ (where the action of $z$ may be different for different fermion species), and ensure that only states with zero global center charge contribute to physical observables. Since $w\lrs{\vec{A}}$ is invariant under center transformations, statistical weight of $z$ is determined by the fermionic action alone.

The representation (\ref{eq:GE_static}) allows to invert the Green-Kubo relations (\ref{eq:GK0}) \emph{analytically} for each static gauge field $\vec{A}\lr{\vec{x}}$ and center element $\Omega = z$, yielding explicit expressions for the spectral function and the corresponding retarded correlator:
\begin{eqnarray}
\label{eq:spectral_function_static}
	S^s\lr{w} = w/T \, \delta\lr{w} \, \vev{\bar{O}^2} 
	+ \nonumber \\ +
	\vev{
		\sum\limits_{m,n}
		\abs{O_{mn}}^2
		\lr{n_z\lr{\epsilon_m} - n_z\lr{\epsilon_n}	}
		\delta\lr{w - \epsilon_n + \epsilon_m }
	} ,
    \\
\label{eq:retarded_static}
	G_R^s\lr{t} = i \theta\lr{t} \vev{ \tr\lr{
        \lrs{n_z\lr{h} O n_z\lr{h}, O\lr{t}} e^{-\beta h} z}} ,
\end{eqnarray}
where $O_{mn} = \bra{\psi_m} O \ket{\psi_n}$, $\lrc{\epsilon_n, \ket{\psi_n}}$ is the eigensystem of $h\lrs{\vec{A}}$, $O\lr{t} = e^{i h t} O e^{-i h t}$, and $\theta\lr{t}$ is the unit step. 

Since the dimensionality of the single-particle fermionic Hilbert space is proportional to spatial lattice volume, the static approximation (\ref{eq:spectral_function_static}) to the full spectral function is calculable with polynomial complexity. We hence propose the following QMC-based spectral reconstruction algorithm:
\begin{enumerate}
	\item Perform the static projection for each Euclidean-time gauge field configuration, which is equivalent to integrating out higher Matsubara modes in the path integral.
	\item Project all Polyakov loops to a single center element $z$ that is closest to their spatial average. To this end, set all time-like links on one of the time slices to $z$, and to $1$ on all other time slices.
	\item Calculate $S^s\lr{w}$ either from the Fourier transform of $G_R^s\lr{t}$ or as a histogram of the eigenvalue differences $w = \epsilon_n - \epsilon_m$ with weights $\abs{O_{mn}}^2 \lr{n_z\lr{\epsilon_n} - n_z\lr{\epsilon_m} }$, if the matrix $h$ is small enough for exact diagonalization.
	\item Refine the static approximation by applying a generic spectral reconstruction method to the difference between $G_E\lr{\tau}$ and $G_E^s\lr{\tau}$. Here we use the MaxEnt method with $S^s\lr{w}$ given by (\ref{eq:spectral_function_static}) as a model function (prior) for this purpose.
\end{enumerate}

A physical interpretation of the static approximation (\ref{eq:spectral_function_static}) and (\ref{eq:retarded_static}) is that we consider real-time fermion dynamics in the background of time-independent gauge fields. These fields form a statistical ensemble that represents the initial conditions for the real-time evolution. The quality of the static approximation hence depends on the validity of two assumptions: the diagonal approximation (\ref{eq:density_matrix}) for the initial-state density matrix and the slowness of gauge field evolution as compared to fermion dynamics. The first assumption can be verified directly in Euclidean-time QMC using purely static quantities such as meson screening masses \cite{Karkkainen:1992jh,Datta:hep-lat/9806034}. The second assumption can only be valid up to a characteristic time scale governing the gauge field dynamics ($\sim \lr{g^2\lr{T} T}^{-1}$ for quark-gluon plasma \cite{Yaffe:hep-ph/9709449}), but still allows us to achieve a parametrically better frequency resolution than $\Delta w \sim \pi \, T$ for generic spectral reconstruction methods based on $G_E\lr{\tau}$ only. Fortunately, in QCD both assumptions are justified at high $T$ where the running coupling $g\lr{T}$ becomes very small.


\emph{$1+1$-dimensional $U\lr{1}$ gauge theory with fermions.} To test our approach, we first consider the Hamiltonian of a $U\lr{1}$ lattice gauge theory in $1+1$ dimensions with two spinless fermion flavours $\hpsi_{1,x}$, $\hpsi_{2,x}$ with $U\lr{1}$ charges $q_1 = +1$, $q_2 = -1$:
\begin{eqnarray}
\label{eq:u1_Hamiltonian}
	\HH
	= 
	\sum\limits_x \frac{g^2 \, \EE_x^2}{2} 
	+ 
	\sum\limits_{x,y,f} \hpsid_{f,x} h_{xy}\lrs{q_f \, \htheta} \hpsi_{f,y} ,
	\\
\label{eq:sp_Hamiltonian}
	h_{xy}\lr{\theta} = \kappa \, e^{i \theta_x} \, \delta_{y,x+1} + \kappa \, e^{-i \theta_{x-1}} \, \delta_{y,x-1} ,
\end{eqnarray}
where $x, \, y = 0 \ldots L_s-1$ label lattice sites on one-dimensional spatial lattice with periodic boundary conditions, $\EE_x = -i \partial/\partial \theta_x$ is the electric field operator canonically conjugate to the $U\lr{1}$ angle-valued link operator $\htheta_x$, and $h_{xy}$ is a single-particle Hamiltonian with nearest-neighbour hopping amplitude $\kappa$ (we set $\kappa = 1$). $g^2$ is the coupling which sets the energy scale for electric flux excitations and hence the characteristic timescale $t \sim g^{-2}$ for gauge field dynamics.

The model (\ref{eq:u1_Hamiltonian}) allows for a finite-temperature Euclidean path integral representation in terms of $U\lr{1}$ spatial and time-like link fields $\theta_{\tau, x}$ and $\Omega_{\tau,x}$ with a non-negative weight. We simulate it using the standard Metropolis algorithm and calculate Euclidean-time correlators $G_E\lr{\tau}$ of a fermionic operator $\hdd = \sum_f q_f \, \lr{\hpsid_{f,1} \hpsi_{f,1} - \hpsid_{f,0} \hpsi_{f,0}}$ (charge difference for two adjacent sites) along with the static approximations (\ref{eq:GE_static}) and (\ref{eq:spectral_function_static}) to $G_E\lr{\tau}$ and the spectral function. Static projection is performed by averaging the link variables $\theta_{\tau, x}$ over $\tau$ for each $x$ after imposing the static gauge where all time-like links $\Omega_{\tau,x}$ are $\tau$-independent.

\begin{figure}[h!tpb]
	\includegraphics[width=0.49\textwidth]{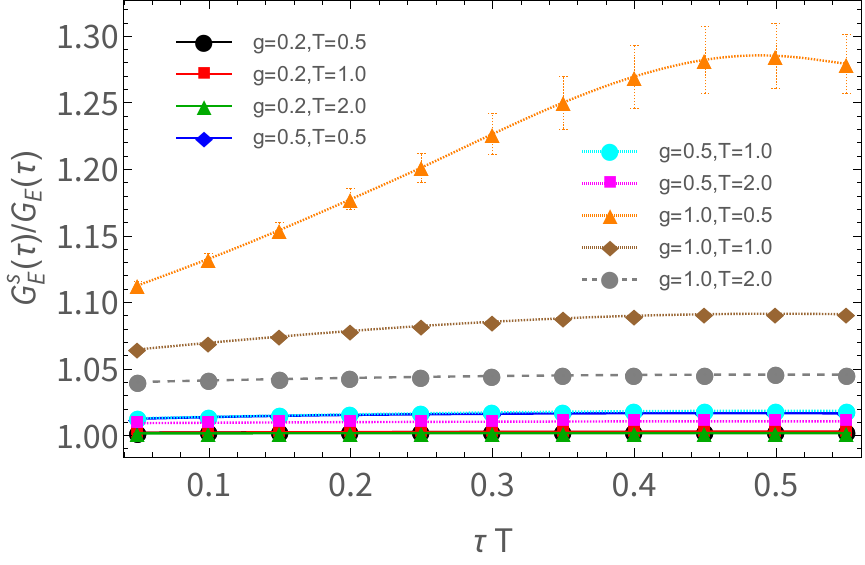}
	\caption{Ratios of Euclidean-time correlators $G_E\lr{\tau}$ of the operator $\hdd = \sum_f q_f \, \lr{\hpsid_{f,1} \hpsi_{f,1} - \hpsid_{f,0} \hpsi_{f,0}}$ and their static approximations $G^s_E\lr{\tau}$ at different couplings $g$ and temperatures $T$ on the lattice with $L_s = 5$.}
	\label{fig:GE_ratios_U1}
\end{figure}

Static approximation to $G_E\lr{\tau}$ becomes increasingly precise for higher temperatures $T$ and smaller couplings $g$, with ratios $G_E^s\lr{\tau}/G_E\lr{\tau}$ quickly becoming $\tau$-independent and approaching unity (see Fig.~\ref{fig:GE_ratios_U1}).

\begin{figure}[h!tpb]
	\includegraphics[width=0.48\textwidth]{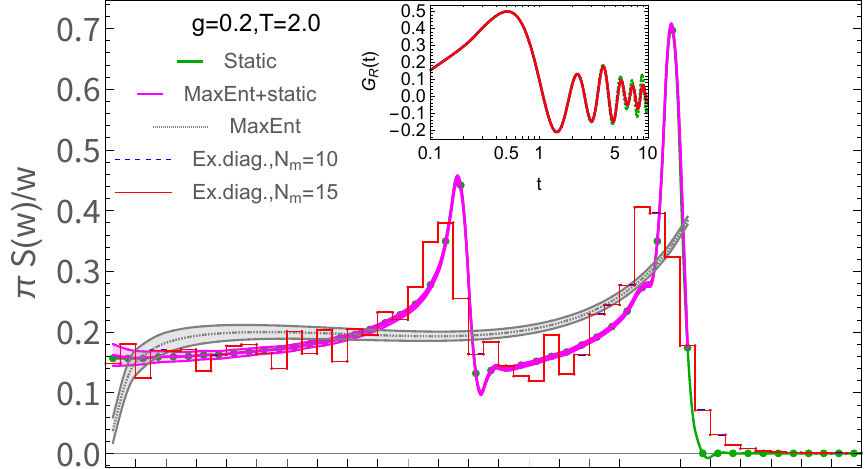}
	\includegraphics[width=0.48\textwidth]{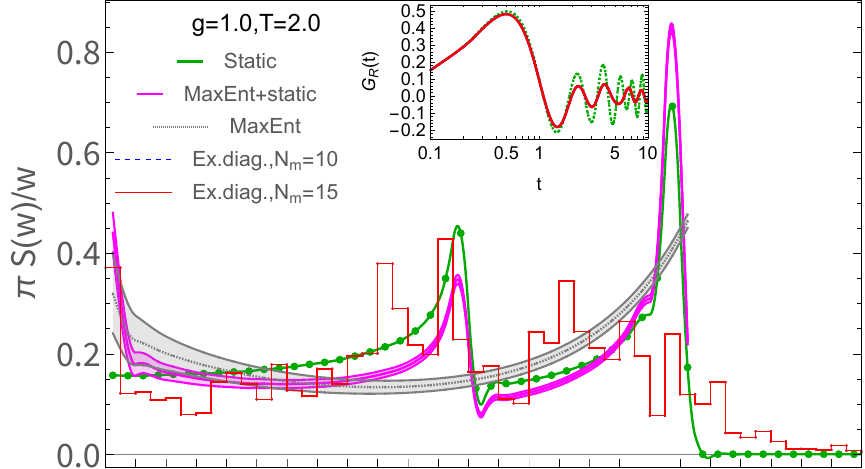}
	\includegraphics[width=0.486\textwidth]{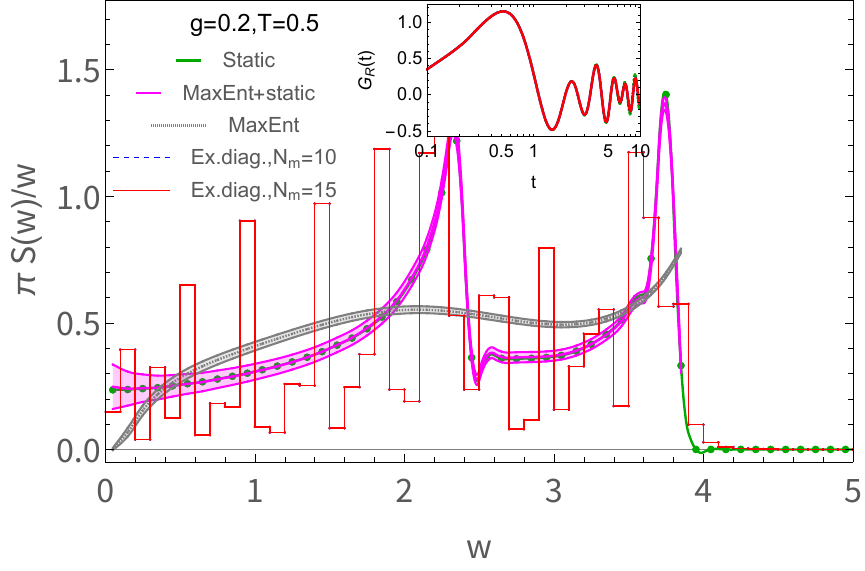}\\
	\caption{Comparison between the full spectral functions $S\lr{w}$ calculated from numerical exact diagonalization of the Hamiltonian (\ref{eq:u1_Hamiltonian}) (red line and dashed blue line for different values of electric flux cut-off $N_m$, see SM for algorithm details) and their static approximations $S^s\lr{w}$ calculated in QMC simulations (green line) using (\ref{eq:spectral_function_static}). We also show numerical estimates of $S\lr{w}$ obtained from MaxEnt method with a constant model function (grey band/line), and with $S^s\lr{w}$ as a model function (magenta band/line). The inset shows retarded correlators $G_R\lr{t}$ (red solid lines) and static approximations thereof (green dashed lines).}
	\label{fig:SP_U1}
\end{figure}

As illustrated in Fig.~\ref{fig:SP_U1}, the static approximation $S^s\lr{w}$ to the spectral function $S\lr{w}$ is very precise in the high-temperature, weak-coupling regime ($g = 0.2$, $T=2.0$), but misses some additional peak structures that appear at either stronger couplings ($g=1.0$, $T = 2.0$) or lower temperatures ($g = 0.2$, $T = 0.5$). Static retarted correlators $G_R^s\lr{t}$ agree with their fully quantum counterparts $G_R\lr{t}$  up to evolution times that become longer as $g$ decreases. These times are parametrically larger than the timescales $t \lesssim \lr{\pi \, T}^{-1}$ probed by Euclidean-time correlators, thus allowing to estimate $S\lr{w}$ with improved frequency resolution. However, larger deviations between $S\lr{w}$ and $S^s\lr{w}$ for $g = 0.2$, $T = 0.5$ as compared with the $g = 0.2$, $T=2.0$ case demonstrate that high temperatures are also important, as they justify the approximation (\ref{eq:density_matrix}) for initial conditions for real-time dynamics.

In contrast, the MaxEnt method \cite{Asakawa:hep-lat/0011040} with $G_E\lr{\tau}$ as input and a standard choice of a constant model function produces only a very smooth estimate of $S\lr{w}$ and fails to reproduce the peak at $w \approx 2.2$ and the low-frequency limit of $S\lr{w}$. Using  the static approximation $S^s\lr{w}$ to $S\lr{w}$ as a model function leads to significantly better results even away from the weak-coupling regime. In particular, the narrow peak in $S\lr{w}$ around $w=0$ at $g=1.0$, $T=2.0$ is reproduced very well, even though it is absent in $S^s\lr{w}$. At lower $T$ the results are less precise, but most prominent peaks of $S\lr{w}$ are still reproduced better than with the constant model function. 


\emph{High-temperature QCD.} We now demonstrate the feasibility of our approach for QCD in the high-temperature deconfined regime. We use \texttt{Gen2L} anisotropic gauge field ensembles generated by the \texttt{FASTSUM} collaboration at temperatures $T = \lrc{760, \, 507 , \, 380} \MeV \gg T_c = 167 \MeV$ ($L_s^3 \times L_t = 32^3 \times \lrc{8,12,16}$ lattices) \cite{FASTSUMGen2L,Glesaaen:2007.04188,Aarts:2209.14681}. Static projection uses a generalization of the ``averaging'' of $SU\lr{3}$-valued spatial links $U_{\tau, \vec{x},i}$ over $\tau$ based on geodesic distance minimization on $SU\lr{3}$ manifold. 

\begin{figure}[h!tpb]
	\includegraphics[width=0.48\textwidth]{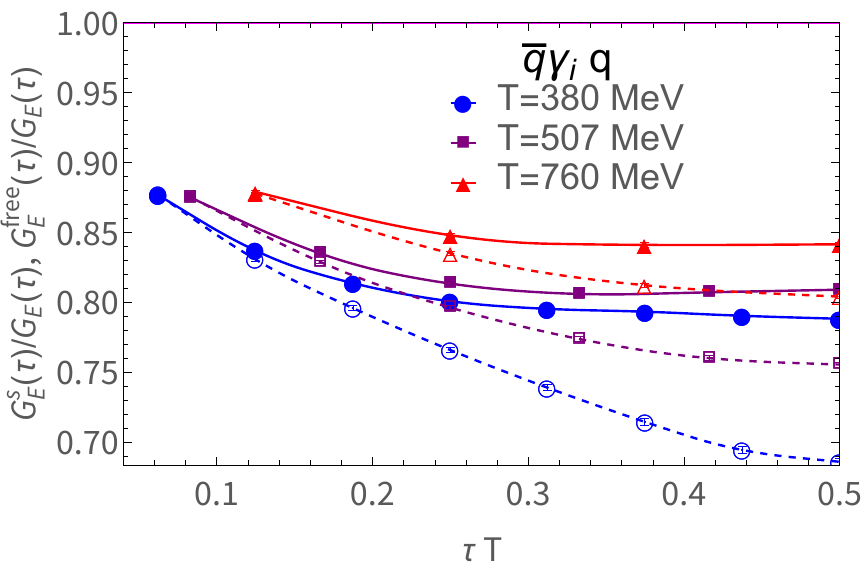}\\
	\caption{Ratios of Euclidean vector-vector correlators calculated on full and static gauge field configurations at different temperatures. For comparison, dashed lines show ratios between free quark correlators and the full lattice QCD data.}
	\label{fig:GE_ratios}
\end{figure}

To assess the quality of the static approximation, we use the \texttt{openqcd-hadspec} code \cite{OpenQCDhadspec} to calculate the Euclidean-time correlators $G_E\lr{\tau}$ and $G_E^s\lr{\tau}$ of local (non-conserved) vector currents $\hat{J}_i = \sum_{\vec{x}}\hat{q}^{\dagger}_{\vec{x}} \gamma_0 \gamma_i \hat{q}_{\vec{x}}$ (equivalently, $\bar{q} \gamma_i q$) for $u$- and $d$-quarks on \texttt{FASTSUM} \texttt{Gen2L} configurations with and without the static projection. Fig.~\ref{fig:GE_ratios} shows that these ratios become almost flat at intermediate values of the Euclidean time $\tau$, which suggests that the static approximation describes the full lattice QCD data quite well (and better than the free quark data) up to contact terms at small $\tau$ and an overall change in vector current renormalization. We note, however, that this statement is observable-dependent, and for some meson correlators $G_E^s\lr{\tau}$ approaches $G_E\lr{\tau}$ slower than the free quark correlators.

We further use the Wilson-Dirac Hamiltonian as a single-particle Hamiltonian $h$ to calculate the static approximation (\ref{eq:spectral_function_static}) to the spectral function in the vector channel. Since the matrix $h$ is now too large for exact diagonalization, we calculate $S^s\lr{w}$ from the Fourier transform of the retarded fermionic correlator (\ref{eq:retarded_static}). We found that the trace over the single-particle fermionic Hilbert space in (\ref{eq:retarded_static}) can be efficiently approximated with $O\lr{10}$ $\mathbb{Z}_4$-valued stochastic estimators \cite{Borsanyi:0809.4711}, which hints at quantum typicality \cite{Bartsch:0902.0927} of quark real-time evolution. Fermi distribution functions $n_z\lr{h}$ in (\ref{eq:retarded_static}) are approximated as minmax polynomials of $h$. With an optimal choice of time evolution method, this allows to calculate $S^s\lr{w}$ with $O\lr{L_s^3}$ computational cost.

\begin{figure}[h!tpb]
	\includegraphics[width=0.48\textwidth]{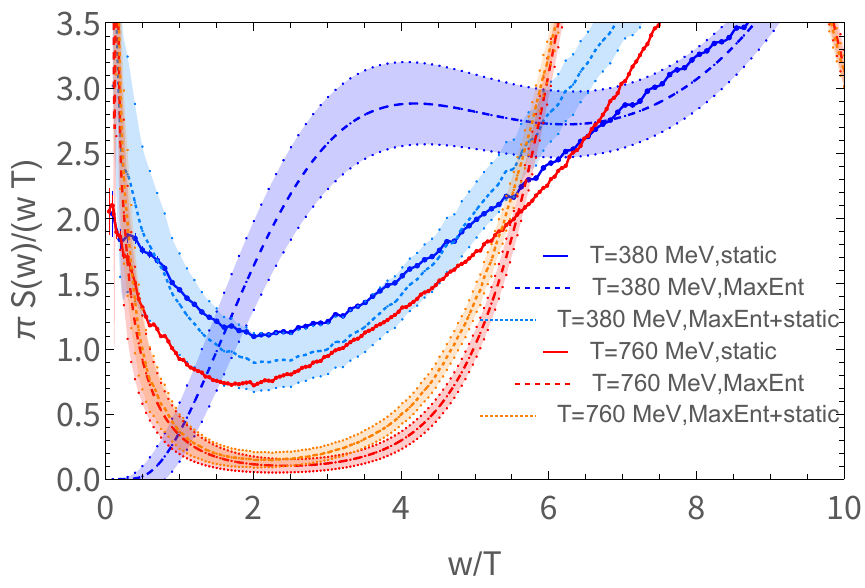}
	\caption{Static approximation $S^s\lr{w}$ to the QCD spectral function in the vector meson channel. We also show the output of the MaxEnt method on the full lattice QCD data with $S^s\lr{w}$ and a constant function as a model function (prior).}
	\label{fig:SP_QCD}
\end{figure}

Plots of $S^s\lr{w}$ in Fig.~\ref{fig:SP_QCD} reveal expected features such as a rising profile at $w \gtrsim 2 T$ and a transport peak at $w \lesssim 2 T$ \cite{Brandt:1212.4200,Aarts:1412.6411}. Even though the difference between the full Euclidean correlator $G_E\lr{\tau}$ and its static approximation is much larger than QMC statistical errors (see Fig.~\ref{fig:GE_ratios}), numerical estimates of $S\lr{w}$ shown on Fig.~\ref{fig:SP_QCD} for $T = 380 \MeV$ ($L_t=16$) demonstrate that the MaxEnt method with $S^s\lr{w}$ as a model function is able to match the full data for $G_E\lr{\tau}$ within statistical errors with relatively small modifications of the low-frequency structure of $S^s\lr{w}$. In contrast, MaxEnt method with a constant model function produces a very different result where the excess of spectral density is concentrated at finite frequencies $w \approx 3 \, T$ rather than in the transport peak. For $T = 760 \MeV$ the MaxEnt method operates on $L_t/2=4$ data points only and loses its expressivity, producing only very narrow peaks around $w = 0$ regardless of the model function. In this regime, physics-informed fits of the difference between $G_E\lr{\tau}$ and $G_E^s\lr{\tau}$ can be more efficient than MaxEnt \cite{Brandt:1212.4200,Buividovich:24:3}.

\emph{Conclusions.} Our method directly incorporates semiclassical approximation to real-time dynamics into the standard Monte-Carlo-based lattice QCD workflow, and allows for reconstruction of spectral functions for fermionic observables with parametrically better frequency resolution than methods based on Euclidean correlators only. The method inherits the exact scale setting and the non-perturbative nature of soft chromo-magnetic fields at high temperatures from QMC, and avoids the technically challenging and unstable classical gauge field evolution. Our approach can be further refined by applying generic spectral reconstruction methods to the small difference between full Euclidean correlators and the static approximations thereof. Beyond spectral reconstruction, fermionic single-particle Hamiltonian can be used to address questions that appear most natural in the Hamiltonian formalism, such as eigenstate thermalization hypothesis and entanglement. It would also be interesting to extend the method to hard gluons (higher Matsubara modes of gauge fields) propagating in static chromo-magnetic field background. This should allow for improved lattice measurements of bosonic spectral functions and the related transport coefficients such as viscosity.

\begin{acknowledgments}
The authors are grateful to members of the \texttt{FASTSUM} collaboration (in particular, Gert Aarts, Chris Allton, Ryan Bignell, Simon Hands, Jon-Ivar~Skullerud, Antonio~Smecca), and to V.~Braguta for many insightful discussions and help with \texttt{FASTSUM} configuration ensembles and codes. P.B. is grateful to A.~Troisi, J.~Ostmeyer and S.~Fratini for many discussions on static approximation in condensed matter physics that were essential for this work. Numerical simulations were undertaken on Barkla, part of the High Performance Computing facilities at the University of Liverpool, UK. This work was partly supported by the STFC Consolidated Grant ST/X000699/1.
\end{acknowledgments}


\begin{widetext}
\section*{Supplemental material: Spectral reconstruction based on dimensional reduction in high-temperature gauge theories}

\section{Quantum Monte-Carlo and numerical exact diagonalization for the \texorpdfstring{${1+1}$}{1+1}-dimensional \texorpdfstring{$U\lr{1}$}{U(1)} lattice gauge theory with fermions}
	\label{sec:u1_lgt}
	
	The Hamiltonian of the ${1+1}$-dimensional $U\lr{1}$ lattice gauge theory with fermions is given and the corresponding notation is explained in the main text. Here we discuss the path integral formulation of this model with Villain-type action, as well as the details of our QMC simulations and exact numerical diagonalization. Simulation code that performs both the QMC simulations and the numerical exact diagonalization is available at \cite{Buividovich:u1_2d_LGT}.
	
	\subsection{Path integral formulation, Villain action and Quantum Monte-Carlo}
	\label{subsec:u1_lgt:villain}
	
	To derive a path integral representation of the partition function $\ZZ = \tr e^{-\beta \HH}$, we use the standard Trotterization prescription and split the Euclidean-time transfer matrix $e^{-\dtau \HH}$ into a product of exponentials of the kinetic energy term and the fermionic term:
	\begin{eqnarray}
	\label{eq:u1_path_int1}
		\ZZ = \tr\lr{\prod\limits_{\tau = 0}^{L_t - 1} e^{-\dtau \HH} }
		= 
		\tr\lr{\prod\limits_{\tau = 0}^{L_t - 1}  
			\expa{-\dtau \sum\limits_{x,y,f} \hpsid_{f,x} h_{xy}\lrs{q_f \, \htheta} \hpsi_{f,y} }
			\expa{-\frac{\dtau \, g^2}{2} \sum\limits_x \EE^2_x} 
		}
		+
		\mathcal{O}\lr{\dtau} ,
	\end{eqnarray}	
	where the trace is over the physical Hilbert space of the theory, that is, states that are invariant under gauge transformations generated by operators of the form
	\begin{eqnarray}
	\label{eq:gauge_transformation_operator}
		\hat{\Omega} = 
        \expa{i \sum\limits_x \Omega_x \lr{\sum\limits_f q_f \, \hpsid_{x,f} \hpsi_{x,f} - \lr{\EE_x - \EE_{x-1}}}}
        =
        \expa{i \sum\limits_{x,f} q_f \, \Omega_x \hpsid_{x,f} \hpsi_{x,f}} \, \expa{ - i \sum\limits_x \Omega_x \lr{\EE_x - \EE_{x-1}}}
	\end{eqnarray}
	for any set of site-dependent gauge transformations with angle-valued parameters $\Omega_x$. The subscript $x-1$ in $\EE_{x-1}$ assumes periodic boundary conditions, so that the lattice site that precedes the site with $x = 0$ is $x = L_s - 1$. The last equality in (\ref{eq:gauge_transformation_operator}) takes into account the commutativity of fermionic and electric field operators.
	
	We now interleave the factors in the last expression in (\ref{eq:u1_path_int1}) with the decomposition of the physical subspace projection operator in the bosonic Hilbert space in terms of the simultaneous eigenstates $\ket{\theta}$ of all angle-valued link operators $\htheta_x$:
	\begin{eqnarray}
	\label{eq:identity_decomposition}
		\hat{\mathcal{P}}
		= 
		\int\limits_{-\pi}^{\pi} \mathcal{D}\theta_x
		\int\limits_{-\pi}^{\pi} \mathcal{D}\Omega_x 
		\ket{\theta_x} \bra{\theta_x} \hat{\Omega} ,
	\end{eqnarray}
	where explicit inclusion of the gauge transformation operator ensures that only gauge singlet states contribute to $\hat{\mathcal{P}}$. Since we integrate over independent $\theta$ and $\Omega$ variables for projectors $\hat{\mathcal{P}}$ inserted between each of the factors in (\ref{eq:u1_path_int1}), we represent $\ZZ$ as a Euclidean-time path integral over angle-valued spatial link fields $\theta_{\tau,x}$ and gauge transformation parameters $\Omega_{\tau,x}$ that depend on the Euclidean time $\tau$ and the spatial lattice site $x$:
	\begin{eqnarray}
	\label{eq:u1_path_int2}
		\ZZ = 
		\int\limits_{-\pi}^{\pi} \mathcal{D}\theta_{\tau,x} \mathcal{D}\Omega_{\tau,x}
		\prod\limits_{\tau = 0}^{L_t - 1}
		\bra{\theta_{\tau}} 
			\expa{-i \sum\limits_x \Omega_{\tau,x} \lr{\EE_x - \EE_{x-1}}} \,
			\expa{-\frac{\dtau \, g^2}{2} \sum\limits_x \EE^2_x }
		\ket{\theta_{\tau+1}}
		\times \nonumber \\ \times
		\tr_F\lr{
		 \prod\limits_{\tau = 0}^{L_t - 1} 
		 \expa{-\dtau \sum\limits_{x,y,f} \hpsid_{f,x} h_{xy}\lrs{q_f \, \theta_{\tau}} \hpsi_{f,y} }
         \,
         \expa{i \sum\limits_{x,f} \Omega_{\tau,x} q_f \, \hpsid_{x,f} \hpsi_{x,f}}
		} ,
	\end{eqnarray}
	where $\tr_F$ denotes trace over the fermionic Hilbert space. The bosonic part $\expa{-i \sum\limits_x \Omega_{\tau,x} \lr{\EE_x - \EE_{x-1}}}$ of the gauge transformation operator (\ref{eq:gauge_transformation_operator}) acts on the link operator eigenstates $\ket{\theta}$ as $ \expa{-i \sum\limits_x \Omega_{\tau,x} \lr{\EE_x - \EE_{x-1}}} \ket{\theta} = \ket{\theta^{\Omega}}$, where $\theta^{\Omega}$ is a gauge-transformed link field:
	\begin{eqnarray}
	\label{eq:gauge_transformation_link}
		\theta^{\Omega}_x = \theta_x + \Omega_{x+1} - \Omega_x .
	\end{eqnarray} 
    We now apply this expression and also use the simultaneous eigenbasis $\ket{n}$ of all electric field operators $\EE_x$ to rewrite each of the factors in the the product over $\tau$ in the first line of (\ref{eq:u1_path_int2}) as
    \begin{eqnarray}
    \label{eq:u1_path_int_factors1}
        \sum\limits_{\lrc{n_x}}
        \braket{\theta_{\tau}^{\Omega}}{n}
			\expa{-\frac{\dtau \, g^2}{2} \sum\limits_{x=0}^{L_s - 1} n^2_x }
		\braket{n}{\theta_{\tau+1}}
        =
        \prod\limits_{x=0}^{L_s - 1}
        \lr{
        \sum\limits_{n_x \in \mathbb{Z}}
        \expa{
            i n_x \lr{\theta_{\tau,x}^{\Omega} - \theta_{\tau+1,x}} - \frac{\dtau \, g^2}{2} n^2_x
            }
        } ,
    \end{eqnarray}
    where the states $\ket{n}$ satisfy $\EE_x \ket{n} = n_x \, \ket{n}$ ,$\braket{\theta_x}{n_x} = \expa{i \sum\limits_x \theta_x \, n_x}$ and are labelled by integer-valued electric fluxes $n_x \in \mathbb{Z}$ through spatial lattice links at each of the lattice sites $x$.

    To arrive at the Villain action that we use for Monte-Carlo sampling of $\theta_x$, we apply the Poisson resummation formula
    \begin{eqnarray}
    \label{eq:u1_poisson_resummation}
        \sum\limits_{n \in \mathbb{Z}} e^{i n \, \theta - \frac{\dtau \, g^2}{2} n^2}
        =
        \sqrt{\frac{2 \pi}{\dtau \, g^2}} \, \sum\limits_{k \in \mathbb{Z}} e^{-\frac{1}{2 \, g^2 \, \dtau} \lr{\theta - 2 \pi \, k}^2}
    \end{eqnarray}
    to each sum over $n_x$ in (\ref{eq:u1_path_int_factors1}). We then obtain the following representation of the path integral:
    \begin{eqnarray}
	\label{eq:u1_path_int3}
		\ZZ = 
		\sum\limits_{k_{\tau,x} \in \mathbb{Z}}
        \int\limits_{-\pi}^{\pi} \mathcal{D}\theta_{\tau,x} \mathcal{D}\Omega_{\tau,x}
		\expa{
         - \frac{1}{2 \, g^2 \, \dtau} \sum\limits_{x, \tau} \lr{\theta_{\tau,x} + \Omega_{\tau,x+1} - \Omega_{\tau,x} - \theta_{\tau+1,x} - 2 \pi \, k_{\tau,x}}^2 
        }
		\times \nonumber \\ \times
		\tr_F\lr{
		 \prod\limits_{\tau = 0}^{L_t - 1} 
		 \expa{-\dtau \sum\limits_{x,y,f} \hpsid_{f,x} h_{xy}\lrs{q_f \, \theta_{\tau}} \hpsi_{f,y} } \,
        \expa{i \sum\limits_{x,f} \Omega_{\tau,x} q_f \, \hpsid_{x,f} \hpsi_{x,f}}  
		} ,
	\end{eqnarray}
    Now the statistical weight associated with link variables $\theta_{\tau,x}$ is manifestly positive, $\Omega_{\tau,x}$ become the temporal link variables, and we have an additional summation over integer-valued variables $k_{\tau,x}$. Since $k_{\tau,x}$ enter on the same footing as the difference $\theta_{\tau,x} + \Omega_{\tau,x+1} - \Omega_{\tau,x} - \theta_{\tau+1,x}$, these variables are in fact associated with plaquettes of the $\lr{1+1}$-dimensional lattice.

    Consider now summation over the integer variables $k_{\tau,x}$ associated with plaquettes of $(1+1)$-dimensional lattice. Let us represent $k_{\tau,x}$ as a sum of two plaquette fields, one of which has a zero total, and another is localized on a single plaquette at $\tau_0, x_0$. To this end, we rewrite 
    \begin{eqnarray}
    \label{eq:u1_hodge1}
     k_{\tau,x} = \lr{k_{\tau,x} - \Phi \delta_{\tau,\tau_0} \delta_{x,x_0}} + \Phi \delta_{\tau,\tau_0} \delta_{x,x_0}
     =
     \bar{k}_{\tau,x} + \Phi \delta_{\tau,\tau_0} \delta_{x,x_0} , 
    \end{eqnarray}
    where $\Phi = \sum\limits_{\tau,x} k_{\tau,x}$ is the total ``magnetic flux'' through the lattice, which we concentrate within a single plaquette at $\tau=\tau_0$, $x=x_0$.
    
    The lattice $2$-form $\bar{k}_{\tau,x}$ with vanishing total can be represented as an external derivative of the $1$-form field $m_{\tau,x,i}$ \cite{Guth:PhysRevD.21.2291}:
    \begin{eqnarray}
    \label{eq:hodge_transform}
    	\bar{k}_{\tau,x} 
        = 
        \lr{d m}_{\tau,x} 
        =  
        m_{\tau,x,1} + m_{\tau,x+1,0} - m_{\tau+1,x,1} - m_{\tau,x,0}
    \end{eqnarray}
	Following the standard procedure of defining the Villain action \cite{Guth:PhysRevD.21.2291}, we now replace summation over $k_{\tau,x}$ by summation over $m_{\tau,x,i}$ and the total flux $\Phi$, times an overall change in the normalization of the partition function. As the next step, we redefine the integration variables 
    \begin{eqnarray}
    \label{eq:u1_noncompact_vars}
        \theta_{\tau,x} \rightarrow \theta_{\tau,x} - 2 \pi m_{\tau,x,1},
        \quad
        \Omega_{\tau,x} \rightarrow \Omega_{\tau,x} - 2 \pi m_{\tau,x,0} .
    \end{eqnarray}
    Integration over original angle-valued varialbes and summation over integer multiples of $2 \pi$ in (\ref{eq:u1_noncompact_vars}) are equivalent to integration over the whole real axis. It is important to notice that the fermionic trace in (\ref{eq:u1_path_int3}) is invariant under such a variable redefinition, as both $\theta_{\tau,x}$ and $\Omega_{\tau,x}$ enter via complex exponentials that are insensitive to shifts by multiples of $2 \pi$.

    Finally, the trace over the fermionic Hilbert space in (\ref{eq:u1_path_int3}) can be represented in the standard fermionic determinant form \cite{Blankenbecler:PhysRevD.24.2278} in terms of the exponentials of the single-particle Hamiltonian. We hence arrive at the following path integral representation:
     \begin{eqnarray}
    \label{eq:u1_path_int_final}
    	\ZZ = 
    	\sum\limits_{\Phi \in \mathbb{Z}}
    	\int\limits_{-\infty}^{+\infty} \mathcal{D}\theta_{\tau,x} \mathcal{D}\Omega_{\tau,x}
    	\expa{
    		- \frac{1}{2 \, g^2 \, \dtau} \sum\limits_{x, \tau} \lr{\theta_{\tau,x} + \Omega_{\tau,x+1} - \theta_{\tau+1,x} - \Omega_{\tau,x} - 2 \pi \, \Phi \delta_{\tau,\tau_0} \delta_{x,x_0}}^2 
        }
    	\times \nonumber \\ \times
    	\prod\limits_f 
        \det\lr{
    		1 + 
    		\prod\limits_{\tau = 0}^{L_t - 1}
    		T_{f,\tau}
    	} ,
    \end{eqnarray}
    where $T_{f,\tau} = e^{-\dtau \, h\lr{q_f \, \theta_{\tau}}} \, e^{i q_f \, \Omega_{\tau}}$ is the fermionic transfer matrix for a single fermion flavour with $U\lr{1}$ charge $q_f$. It includes both the matrix exponential of the single-particle Hamiltonian for time slice $\tau$ and the time-like links $e^{i q_f \Omega_{\tau}}$ which act on the single-particle fermionic Hilbert space as a diagonal matrix, that is, $\lr{e^{i q_f \Omega_{\tau}} \phi }_x = e^{i q_f \Omega_{\tau, x}} \phi_x$.

    As usual, the trace over the Hilbert subspace associated with each of the fermion flavours gives a single determinant for each flavour. For oppositely charged fermion species with otherwise equal parameters the two determinants are complex-conjugate to each other, so their product gives the absolute value squared of one of the determinants, and hence a manifestly positive path integral weight.

    Finally, we notice that the path integral (\ref{eq:u1_path_int_final}) is invariant under gauge transformations that depend on $\tau$ and $x$:
    \begin{eqnarray}
    \label{eq:u1_gauge_transformations}
        \theta_{\tau,x}^{\omega} = \theta_{\tau,x} + \omega_{\tau,x+1} - \omega_{\tau,x} ,
        \quad
        \Omega_{\tau,x}^{\omega} = \Omega_{\tau,x} + \omega_{\tau+1,x} - \omega_{\tau,x} ,
    \end{eqnarray}
    where $\omega_{\tau,x}$ can take arbitrary real values.
    
    Our motivation to use the Villain action and the non-local fermionic action that involves exponentials $e^{-\dtau \, h}$ of the single-particle Hamiltonian is that such a formulation has extremely small time discretization artifacts and allows for very precise comparison with the results of numerical exact diagonalization.

    \subsection{Monte-Carlo sampling}
    \label{subsec:u1_lgt:monte_carlo}

    Since the integration variables $\theta_{\tau,x}$ and $\Omega_{\tau,x}$ in (\ref{eq:u1_path_int_final}) can take arbitrary values, these gauge transformations parametrize flat directions of the Villain action. In contrast to Monte-Carlo integration over compact variables, in this case one needs gauge fixing to make the path integral (\ref{eq:u1_path_int_final}) well-defined. For our Monte-Carlo sampling we use the gauge where $\Omega_{\tau,x} = 0$ for all $\tau = 0 \ldots L_t - 2$, and takes nonzero values at the time slice $\tau = L_t - 1$.
    
    We generate each new gauge field configuration by first generating a random flux number $\Phi$ according to the probability distribution of $\Phi$ for pure gauge theory, that can be easily obtained by taking the Gaussian integral over $\theta_{\tau,x}$ and $\Omega_{\tau,x}$ in the partition function (\ref{eq:u1_path_int_final}) at fixed $\Phi$:
    \begin{eqnarray}
    \label{eq:u1_flux_probability}
        p\lr{\Phi} 
        = 
        \mathcal{N} \, \expa{-\frac{\lr{2 \pi}^2 \, \Phi^2}{2 \, g^2 \, \dtau \, L_t \, L_s}}
        =
        \mathcal{N} \, \expa{-\frac{\lr{2 \pi}^2 \, \Phi^2}{2 \, g^2 \, \beta \, L_s}}
    \end{eqnarray}
    After generating $\Phi$, we generate configurations of $\Omega_{\tau,x}$ and $\theta_{\tau,x}$ from the multi-dimensional Gaussian distribution given by the first exponential factor in (\ref{eq:u1_path_int_final}), which is easily reduced to a product of independent one-dimensional Gaussian distributions by using the Fourier transform over $\tau$ and $x$.  

    After this, we perform the Metropolis accept-reject step that includes the squared absolute value of the fermionic determinant in the second line of (\ref{eq:u1_path_int_final}), thus recovering the correct path integral weight. For small lattice sizes and high temperatures, the fermionic determinant is numerically well-behaved, and we get reasonable acceptance probabilities between $0.1$ and $0.9$ depending on the parameters of the Hamiltonian.

    \subsection{Static gauge and static projection for \texorpdfstring{$U\lr{1}$}{U(1)} gauge fields}
    \label{subsec:u1_lgt:static}

    Here we justify the static projection procedure that we explain in the main text. First of all we note that according to (\ref{eq:u1_flux_probability}), nonzero values of the ``total magnetic flux'' $\Phi$ are strongly suppressed in the limit of small $g$ and large temperatures, that is, exactly in the limit when static approximation becomes asymptotically exact. Let us therefore consider the bosonic part of the action of our $U\lr{1}$ lattice gauge theory (first line in (\ref{eq:u1_path_int_final})) with $\Phi = 0$. Separating the finite differences $\theta_{\tau+1,x} - \theta_{\tau,x}$ and $\Omega_{\tau,x+1} - \Omega_{\tau,x}$, we can rewrite this part of the action as
    \begin{eqnarray}
    \label{eq:u1_gauge_action1}
    \frac{1}{2 \, g^2 \, \dtau} \sum\limits_{x, \tau} \lr{\theta_{\tau,x} + \Omega_{\tau,x+1} - \theta_{\tau+1,x} - \Omega_{\tau,x}}^2
    = \nonumber \\ = 
    \frac{1}{2 \, g^2 \, \dtau} \sum\limits_{x, \tau}
    \lr{\theta_{\tau+1,x} - \theta_{\tau,x}}^2
    +
    \frac{1}{2 \, g^2 \, \dtau} \sum\limits_{x, \tau}
    \lr{\Omega_{\tau,x+1} - \Omega_{\tau,x}}^2
    -
    \frac{1}{g^2 \, \dtau} \sum\limits_{x, \tau}
    \lr{\theta_{\tau+1,x} - \theta_{\tau,x}}
    \lr{\Omega_{\tau,x+1} - \Omega_{\tau,x}} .
    \end{eqnarray}
    The last term in the above equation couples time-like and space-like link variables, and can be represented in several equivalent forms (taking into account the periodic boundary conditions):
    \begin{eqnarray}
    \label{eq:u1_cross_term}
        \frac{1}{g^2 \, \dtau} \sum\limits_{x, \tau}
    \lr{\theta_{\tau+1,x} - \theta_{\tau,x}}
    \lr{\Omega_{\tau,x+1} - \Omega_{\tau,x}}
    = \nonumber \\ =
    \frac{1}{g^2 \, \dtau} \sum\limits_{x, \tau}
    \theta_{\tau,x}
    \lr{
    \Omega_{\tau-1,x+1} - \Omega_{\tau-1,x}
    -
    \Omega_{\tau,x+1} + \Omega_{\tau,x}
    }
    = \nonumber \\ = 
    \frac{1}{g^2 \, \dtau} \sum\limits_{x, \tau}
    \Omega_{\tau,x}
    \lr{
    \theta_{\tau+1,x-1} - \theta_{\tau,x-1}
    -
    \theta_{\tau+1,x} + \theta_{\tau,x}
    } .
    \end{eqnarray}
    These representations make it clear that the cross-product term (last summand in (\ref{eq:u1_gauge_action1})) can be made to vanish, and the action can be split into independent actions for space-like and time-like links in two cases: either $\Omega_{\tau,x}$ is $\tau$-independent (and hence the combination of $\Omega$'s inside the brackets in the second line of (\ref{eq:u1_cross_term}) vanishes), or $\theta_{x,\tau}$ is $x$-independent (and hence the combination of $\theta$'s inside the brackets in the third line of (\ref{eq:u1_cross_term}) vanishes). The latter choice is equivalent to Coulomb gauge for our $\lr{1 +1}$-dimensional $U\lr{1}$ lattice gauge theory. In this work, however, we consider the second option with $\tau$-independent time-like link variables $\Omega_{\tau,x}$, which we refer to as \emph{static gauge}.
    
    Static gauge can be imposed for any configuration of spatial and temporal link variables $\theta_{\tau,x}$ and $\Omega_{\tau,x}$. To this end we can apply a gauge transformation that makes the temporal links $\Omega_{\tau,x}$ $\tau$-independent and equal to $\bar{\Omega}_x = L_t^{-1} \sum\limits_{\tau} \Omega_{\tau,x}$. Using (\ref{eq:u1_gauge_transformations}), such a gauge transformation $\omega_{\tau,x}$ can be explicitly constructed for each spatial site $x$ from the equation:
    \begin{eqnarray}
    \label{eq:u1_static_gauge}
        \Omega^{\omega}_{\tau,x}
        =
        \Omega_{\tau,x} + \omega_{\tau+1,x} - \omega_{\tau,x}
        =
     \bar{\Omega}_{\tau,x}
     \Rightarrow
     \omega_{\tau+1,x} = \omega_{\tau,x} + \lr{\bar{\Omega}_x - \Omega_{\tau,x}} .
    \end{eqnarray}
    These equations can be solved for all $\tau$ starting with e.g. $\omega_{0,x} = 0$.  

    Imposing the static gauge, we can rewrite the gauge action (\ref{eq:u1_gauge_action1}) as
    \begin{eqnarray}
    \label{eq:u1_static_gauge_action}
    \frac{1}{2 \, g^2 \, \dtau} \sum\limits_{x, \tau} \lr{\theta_{\tau,x} + \Omega_{\tau,x+1} - \theta_{\tau+1,x} - \Omega_{\tau,x}}^2
    = \nonumber \\ = 
    \frac{1}{2 \, g^2 \, \dtau} \sum\limits_{x, \tau}
    \lr{\theta_{\tau+1,x} - \theta_{\tau,x}}^2
    +
    \frac{L_t}{2 \, g^2 \, \dtau} \sum\limits_{x}
    \lr{\bar{\Omega}_{x+1} - \bar{\Omega}_{x}}^2   .
    \end{eqnarray}
    This representation makes it clear that $\tau$-independent spatial link variables $\theta_{\tau,x}$ correspond to the global minimum of the Villain action (\ref{eq:u1_gauge_action1}) at $\Phi = 0$, and that the kinetic part of the gauge action does not suppress the $\tau$-independent component of the spatial link field $\theta_{\tau,x}$. In higher-dimensional gauge theories, spatial link fields will still have a nontrivial gauge action associated with spatial plaquettes.
    
    It is also instructive to express this action in terms of the non-compact Polyakov loop variables $P_x = \sum\limits_{\tau} \Omega_{\tau,x} = L_t \bar{\Omega}_x$ (technically, they are the logs of the $U\lr{1}$-valued Polyakov loops plus some multiples of $2 \pi$). The second summand in (\ref{eq:u1_static_gauge}) is then expressed as $\frac{L_t}{2 \, g^2 \, \dtau L_t^2} \sum\limits_{x}
    \lr{P_{x+1} - P_{x}}^2 = \frac{T}{2 \, g^2} \sum\limits_{x}
    \lr{P_{x+1} - P_{x}}^2$ . This representation makes it clear that at the perturbative level spatial fluctuations of Polyakov loops are suppressed by the same power of temperature as higher Matsubara components of space-like gauge fields. Therefore, to arrive at a consistent static approximation, we have to neglect spatial dependence of Polyakov loops along with the Euclidean time dependence of space-like gauge fields. Only the $x$-independent component $\bar{P} = L_s^{-1} \sum\limits_x P_x$ of the Polyakov loop is not suppressed. The $U\lr{1}$ phase $e^{i \bar{P}}$ becomes the global center variable $z$ in the expressions for the static approximation to the Euclidean and real-time correlators and the spectral function in the main text.

    \subsection{Euclidean-time and real-time correlators of fermionic operators}
    \label{subsec:u1:correlators}

    Explicit expressions for correlators of fermionic bilinear operators such as $\hat{\Delta} = \hat{q}_1 - \hat{q}_0$ are most conveniently written in terms of Fermionic Green's functions for each flavour $f$, which we write down as matrices of the same dimension as the fermionic single-particle Hamiltonian $h\lrs{\theta_x}$: 
    \begin{eqnarray}
    \label{eq:u1:fermionic_Green}
     G_{f,\tau_1, \tau_2}
     =
     \begin{cases}
        \lr{1 + \prod\limits_{\tau=\tau_1}^{L_t-1} T_{f,\tau} \, \prod\limits_{\tau=0}^{\tau_1-1} T_{f,\tau} }^{-1} \, \prod\limits_{\tau=\tau_1}^{\tau_2-1} T_{f,\tau}
         , & \tau_2 > \tau_1 \\
        \lr{1 + \prod\limits_{\tau=\tau_1}^{L_t-1} T_{f,\tau} \, \prod\limits_{\tau=0}^{\tau_1-1} T_{f,\tau} }^{-1} \, 
         , & \tau_2 = \tau_1 \\
        \prod\limits_{\tau=\tau_2}^{L_t-1} T_{f,\tau}
        \,
        \lr{
        1 + \prod\limits_{\tau=0}^{L_t-1} T_{f,\tau}
        }^{-1}
        \,
        \prod\limits_{\tau=0}^{\tau_1-1} T_{f,\tau} ,
        & \tau_2 < \tau_1 \\
     \end{cases}  ,
    \end{eqnarray}
    where we assume that both $\tau_1$ and $\tau_2$ take values in the range $0 \ldots L_t - 1$. 
    
    Introducing the single-particle operator $\lr{\Delta_f}_{x,y} = q_f \lr{\delta_{x,1} \delta_{y,1} - \delta_{x,0} \delta_{y,0}}$ that corresponds to the operator $\hat{\Delta}$, Euclidean-time fermionic correlator $G_E\lr{\tau} = \ZZ^{-1} \tr\lr{\hat{\Delta} e^{-\tau \HH} \hat{\Delta} e^{-\lr{\beta - \tau} \HH}}$ can be calculated as
    \begin{eqnarray}
    \label{eq:u1:fermionic_corr1}
        G_E\lr{\tau}
        =
        \vev{
        \sum\limits_f 
        \tr\lr{
        \Delta_f \, G_{f,0,\tau} \, \Delta_f \, G_{f,\tau, 0} }
        +
        \sum\limits_{f,f'}\tr\lr{
        \Delta_f \, G_{f,0,0}} \, \tr\lr{ \Delta_{f'} \, G_{f',\tau, \tau} } 
        },
    \end{eqnarray}
    where the first and the second contributions correspond to connected and disconnected fermionic diagrams, respectively, and the trace goes over the direct sum of Hilbert spaces that correspond to both flavours. The expectation value is taken over the ensemble of gauge field configurations (spatial links $\theta_{\tau,x}$ and time-like links $\Omega_{\tau,x}$) with the weight given by the integrand in (\ref{eq:u1_path_int_final}). We used $2 \cdot 10^5$ statistically independent field configurations in our analysis.
    
    It is also instructive to express this operator in terms of single-flavour Hamiltonian with $q_f = +1$, taking into account that the change of the $U\lr{1}$ charge sign between the two flavours is equivalent to complex conjugation:
    \begin{eqnarray}
    \label{eq:u1:fermionic_corr_final}
        G_E\lr{\tau} = 
        \vev{
        2 \re \tr\lr{
        \Delta \, G_{0,\tau} \, \Delta \, G_{\tau, 0} }
        +
        4 \im \tr\lr{
        \Delta \, G_{0,0}} \, \im \tr\lr{ \Delta \, G_{\tau, \tau} } 
        },
    \end{eqnarray}
    where now all the operators act on a single-particle Hilbert subspace of a single flavour with $q_f = +1$ only.

    Finally, we note that the operator $\hat{\Delta} = \hat{q}_1 - \hat{q}_0$ was chosen because of the nontrivial two-peak structure in the spectral function. In contrast, the spectral function of the electric current operator in this model features a single peak that is very close to the origin and cannot be resolved neither in the static approximation nor in a MEM analysis. The corresponding Euclidean correlator is almost $\tau$-independent. This is probably due to the absence of scattering in this channel for our $\lr{1+1}$-dimensional model.

    \subsection{Numerical exact diagonalization of the full Hamiltonian of \texorpdfstring{$U\lr{1}$}{U(1)} lattice gauge theory with fermions}

    We perform the numerical diagonalization of the full Hamiltonian of our $U\lr{1}$ lattice gauge theory directly in the physical (gauge-invariant) Hilbert space. To construct a suitable basis in this space, we use the eigenbasis $\ket{\lrc{n_{f,x}}}$ of fermion occupation number operators $\hat{n}_{f,x} = \hpsid_{f,x} \hpsi_{f,x}$. For bosonic Hilbert space we use the eigenbasis $\ket{\lrc{E_x}}$ of electric field operators $\EE_x$ with $\EE_y \ket{\lrc{E_x}} = E_y \ket{\lrc{E_x}}$ for any spatial lattice link. Basis state vectors $\ket{\lrc{n_{f,x}}, \lrc{E_x}}$ in the full Hilbert space are hence labelled by the fermion occupation numbers $n_{f,x}$ for all lattice sites and electric field fluxes $E_x$ for all lattice links. In this basis, the $U\lr{1}$ gauge constraint operator is fully diagonal. Its vanishing is equivalent to the Gauss law for electric flux numbers and local $U\lr{1}$ charges:
    \begin{eqnarray}
    \label{eq:u1_gauss_law}
        E_x - E_{x-1} = \sum\limits_f q_f \, n_{f,x} .
    \end{eqnarray}
    For our $U\lr{1}$ lattice gauge theory with a single spatial dimension, the full basis of physical states that satisfy (\ref{eq:u1_gauss_law}) can be constructed in the following way. First, we pick only states with zero total $U\lr{1}$ charge $\hat{Q} = \sum\limits_{f,x} q_f \, \hat{n}_{f,x}$. States with a nonzero total $U\lr{1}$ charge cannot be physical, which can be easily seen by summing the constraint equation (\ref{eq:u1_gauss_law}) over $x$ on a finite lattice with periodic boundary conditions. We then dress these zero-charge states with some ``minimal'' electric fluxes to obtain a set of ``parent'' basis states. Our prescription is to start with $E_0 = 0$ and consecutively assign electric fluxes according as $E_{x+1} = E_x + \sum\limits_f q_f \, n_{f,x+1}$ for $x = 0 \ldots L_s - 1$. The full basis that we use for numerical diagonalization is then obtained by shifting all local fluxes $E_x \rightarrow E_x + E$ by a global $x$-independent flux $E = -N_m \ldots N_m$. Since all $E_x$ are shifted simultaneously, this does not violate the Gauss law (\ref{eq:u1_gauss_law}). All physical states can therefore be uniquely enumerated by a collection of fermionic occupation numbers (with zero charge) and a single integer-valued global flux $E$. The parameter $N_m$ defines our truncation prescription to turn an infinite-dimensional bosonic Hilbert space into a finite Hilbert space that can be treated numerically. Since we assigned a special role to the origin $x = 0$ when constructing our basis, at any finite $N_m$ this basis violates translational symmetry. However, the effects of this violation very quickly become unimportant, as only high-energy states with very small Boltzmann weights are affected. The set of low-energy states is perfectly translationally invariant.

    The fact that the dimensionality of the bosonic Hilbert space does not depend on the spatial lattice size reflects the topological nature of the gauge sector of $1+1$-dimensional gauge theories. Nevertheless, these theories still feature nontrivial dynamical properties in the presence of fermions.

    The Hamiltonian acts on any state $\ket{\lrc{n_{f,x}} \lrc{E_x}}$ as
    \begin{eqnarray}
    \label{eq:u1_hamiltonian_action}
        \HH \ket{\lrc{n_{f,x}} \lrc{E_x}}
        =
        \frac{g^2}{2} \sum\limits_x E_x^2 \ket{\lrc{n_{f,x}} \lrc{E_x}}
        +
        \kappa \,
        \sum\limits_{y,s,f}
        \sigma\lrs{y,s,f,\lrc{n_{f,x}}}
        \ket{\lrc{n_{f,x}'} \lrc{E_x'}}
    \end{eqnarray}
    where the states $\ket{\lrc{n_{f,x}'} \lrc{E_x'}}$ are obtained by moving a fermion with flavour $f$ from lattice site $y$ to an adjoint lattice site $y+s$ in either forward ($s = +1$) or backward ($s = -1$) direction, and adjusting the electric fluxes accordingly. More specifically, if $n_{f,y} = 1$ and $n_{f,y+s} = 0$, we set $n_{f,y}' = 0$ and $n_{f,y+s}' = 1$. To ensure that the Gauss law is satisfied, we also change the fluxes as $E_y' = E_y - q_f$ if $s = +1$, and $E_{y-1}' = E_{y-1} + q_f$ if $s = -1$. All other occupation numbers and fluxes are unchanged for given $y$, $s$ and $f$. $\sigma\lrs{y,s,f,\lrc{n_{f,x}}}$ is the corresponding ``hopping sign'' which ensures that proper fermionic commutation relations are satisfied. Namely, if hopping is not possible because either $n_{f,y} = 0$ (no particle to hop) or $n_{f,y+s} = 1$ (new fermion position is already full), $\sigma\lrs{y,s,f,\lrc{n_{f,x}}} = 0$. With a finite cutoff $E_{max}$ on electric fluxes, $\sigma\lrs{y,s,f,\lrc{n_{f,x}}}$ is also set to zero if a fermion hopping creates too large a flux number that is not contained within our truncated basis. Finally, if hopping is allowed, $\sigma\lrs{y,s,f,\lrc{n_{f,x}}}$ is equal to $\pm 1$ depending on whether we cross an even or odd number of lattice sites occupied by an odd number of fermions with flavour $f$ when dragging a fermion from site $y$ to $y+s$. This number is nontrivial as the shift operation $y + s$ takes into account periodic boundary conditions.

    Once the Hamiltonian is represented as a finite matrix upon imposing a finite electric flux cut-off $N_m$, spectral functions are obtained by histogramming the frequencies $w = E_n - E_m$ with the weight $\abs{\bra{n} O \ket{m}}^2 \lr{e^{-\beta E_m} - e^{-\beta E_n}}$. Finally, convergence with respect to $N_m$ is controlled by comparing the results at different $N_m$. With temperatures $T \lesssim 2$ used in our work, we found that the results for the spectral functions are almost indistinguishable for $N_m = 10$ and $N_m = 15$. We note that exact diagonalization studies of semi-classical high-temperature regime require larger and larger dimensions of the bosonic Hilbert space at higher temperatures, and are hence more technically challenging than exact diagonalization studies at low temperatures. In particular, the popular truncation scheme with $j_{max} = 1/2$ for $SU\lr{2}$ gauge theory \cite{Schaefer:2308.16202} would be inadequate in this regime. 
    
\section{Static projection and fermion real-time evolution in lattice QCD}
\label{sec:lqcd}

In this Section, we discuss the details of our numerical analysis of fermionic correlators in lattice QCD that is based on \texttt{FASTSUM} \texttt{Gen2L} ensembles of gauge field configurations. We use all available gauge field configurations for each value of temperature. The code that reads lattice gauge field configurations in the \texttt{OpenQCD} format and performs static projection and the Hamiltonian-based calculation of fermionic real-time and Euclidean-time correlators in the static approximation is available at \cite{Buividovich:static_lattice_QCD}.

\subsection{Static projection for \texorpdfstring{$SU\lr{3}$}{SU(3)} lattice gauge fields}
\label{subsec:lqcd:static_projection}
	
In a continuum gauge theory, projection of gauge fields to their zero Matsubara frequency component is as simple as integration or summation of spatial gauge fields $\vec{A}\lr{\tau, \vec{x}}$ over $\tau$ for each $\vec{x}$. This prescription also works for angle-valued or non-compact link variables describing $U\lr{1}$ gauge fields. However, in the absence of a natural ``addition'' operation for $SU\lr{N}$-valued link variables, the process of calculating averages over $\tau$ appears less intuitive. Here we describe our generalization of the operation of taking a mean value of several $SU\lr{N}$ group elements which we denote in a general form as $U_k$, $k = 1 \ldots M$. In practice, $U_k$ will be the spatial link variables for a fixed spatial lattice site and all values of Euclidean time $\tau = 0 \ldots L_t - 1$. 

Our generalization is based on the notion of a geodesic distance between two $SU\lr{N}$ group elements $U$ and $V$, which can be written in terms of the corresponding $N \times N$ matrices in the fundamental group representation as $1 - \frac{1}{N} \, \re \tr\lr{U V^{\dag}}$. We then define the ``mean value'' of $U_k \in SU\lr{N}$ as a group element $\bar{U} \in SU\lr{N}$ such that the sum of geodesic distances 
\begin{eqnarray}
\label{eq:su3_mean_def0}
	d^2\lr{\bar{U}} = \sum\limits_{k=1}^M \lr{ 1 - \frac{1}{N} \re \tr\lr{\bar{U} U_k^{\dag}} }
\end{eqnarray}
between $\bar{U}$ and all $U_k$ takes minimal possible value. It is convenient to define an $N \times N$ complex matrix $S = \sum\limits_{k=1}^M U_k$, and rewrite (\ref{eq:su3_mean_def0}) as
\begin{eqnarray}
	\label{eq:su3_mean_def1}
	d^2\lr{\bar{U}} = M - \frac{1}{N} \, \re \tr\lr{\bar{U} S^{\dag}} .
\end{eqnarray}
At any local minimum or maximum of $d^2\lr{\bar{U}}$ as a function of $\bar{U}$, both left and right derivatives of $d^2\lr{\bar{U}}$ over $\bar{U}$ should vanish. Taking for example the left derivative on $SU\lr{N}$ manifold which acts on $\bar{U}$ as $\nabla_a \bar{U} = i T_a \bar{U}$, we conclude that the following equation should hold at any local extremum:
\begin{eqnarray}
\label{eq:su3_extr_eq1}
	\tr\lr{T_a \lr{\bar{U} S^{\dag} - S \bar{U}^{\dag}}} = 0 ,  
\end{eqnarray}
where $T_a$, $a = 1 \ldots N^2 - 1$ are the generators of $SU\lr{N}$. Since $T_a$ form a full basis in the space of hermitian, traceless matrices, the equation (\ref{eq:su3_extr_eq1}) can only hold if
\begin{eqnarray}
\label{eq:su3_extr_eq2}
	\bar{U} S^{\dag} - S \bar{U}^{\dag} = 2 i \xi \, I ,
\end{eqnarray}
where $I$ is the identity matrix and $\xi$ is some real number. Let us now consider the Singular Value Decomposition (SVD) of the matrix $S$: $S = W \Sigma V^{\dag}$, where $W$ and $V$ are some unitary matrices and $\Sigma$ is a diagonal matrix with positive entries $\sigma_j$, $j = 1 \ldots N$. Substituting this SVD into (\ref{eq:su3_extr_eq2}), it is easy to see that the equation (\ref{eq:su3_extr_eq2}) is solved by the ansatz
\begin{eqnarray}
\label{eq:su3_min_ansatz}
	\bar{U} = W A V^{\dag} ,
\end{eqnarray} 
where $A = \mathrm{diag}\lr{e^{i \alpha_j}}$ is a diagonal matrix with entries $e^{i \alpha_j}$, $\alpha_j \in \lrs{-\pi, \pi}$. Substituting both the SVD of $S$ and the ansatz (\ref{eq:su3_min_ansatz}) into (\ref{eq:su3_extr_eq2}), we rewrite this equation as
\begin{eqnarray}
\label{eq:su3_extr_eq3}
    A \Sigma - \Sigma A = 2 i \xi \, I .
\end{eqnarray}
Since $A$, $I$ and $\Sigma$ are all diagonal matrices, this equation implies that the following equalities should hold for each of the diagonal entries:
\begin{eqnarray}
\label{eq:su3_extr_eq4}
    e^{i \alpha_j} \, \sigma_j - \sigma_j \, e^{i \alpha_j} = 2 i \xi ,
\end{eqnarray}
or, equivalently, 
\begin{eqnarray}
\label{eq:su3_extr_eq5}
    \sin\lr{\alpha_j} = \xi/\sigma_j 
    \, \Rightarrow \,
    \alpha_j = \arcsin\lr{\xi/\sigma_j} .
\end{eqnarray}
This equation allows to express the diagonal entries $\alpha_j$ in terms of a single free variable $\xi$, up to different choices of the branch of the $\arcsin$ function. 

We now need to remember about the $SU\lr{N}$ constraint $\det\lr{\bar{U}} = 1$, which implies that
\begin{eqnarray}
\label{eq:su3_det_condition}
    \det\lr{W V^{\dag}} e^{i \sum\limits_j \alpha_j}
    =
    \det\lr{W V^{\dag}} e^{i \sum\limits_j \arcsin\lr{\xi/\sigma_j}}
    = 1 .
\end{eqnarray}
Since $W$, $V^{\dag}$ and $\sigma_j$ are all known from the SVD of the matrix $S$, all we need to do is to solve the above equation in terms of $\xi$, which apparently can only take values between $-\max\limits_j \sigma_j$ and $+\max\limits_j \sigma_j$, so that the argument of the $\arcsin$ function is between $-1$ and $1$.

The equation (\ref{eq:su3_det_condition}) can have several solutions that correspond to different branches of the $\arcsin$ function: $\sin\lr{\phi} = x \Rightarrow \phi = \pi m + \lr{-1}^m \, \arcsin_0\lr{x}$, where $\arcsin_0\lr{x} \in \lrs{-\pi/2 \ldots \pi/2}$ for $x \in \lrs{-1 \ldots 1}$ and $m \in \mathbb{Z}$ is an arbitrary integer. Each of these solutions will correspond to some local extremum. To find the global minimum of the sum of geodesic distances (\ref{eq:su3_mean_def1}), we scan over the branches with $m_j = 0, 1$ of $\arcsin\lr{\xi/\sigma_j}$ for each $j$. Branches with other values of $m_j$ differ from these branches by multiples of $2 \pi$ that yield the same values of $e^{i \alpha_j}$. For each choice of $m_j$, the equation (\ref{eq:su3_det_condition}) is solved using the bisection method, starting with $\xi = -\max\limits_j \sigma_j$ and $\xi = +\max\limits_j \sigma_j$ as interval endpoints. 

This general algorithm is applied to perform static projection for space-like $SU\lr{3}$-valued link variables. We first impose the static gauge where time-like links $U_{\tau,x,0}$ are time-independent. A straightforward generalization of the equation (\ref{eq:u1_static_gauge}) from $U\lr{1}$ to $SU\lr{3}$ gauge group allows to achieve this by applying a gauge transformation $\omega_{\tau,\vec{x}} \in SU\lr{3}$ that satisfies the equation
\begin{eqnarray}
\label{eq:sun_static_gauge}
    U^{\omega}_{\tau,\vec{x},0}
    =
    \omega^{\dag}_{\tau,\vec{x}} U_{\tau,\vec{x},0} \omega_{\tau+1,\vec{x}}
    =
    \bar{\Phi}_{\vec{x}}
    \Rightarrow
    \omega_{\tau+1,\vec{x}} = U^{\dag}_{\tau,\vec{x},0} \omega_{\tau,\vec{x}} \bar{U}_{\vec{x},0} ,
\end{eqnarray}
where $\bar{U}_{\vec{x},0} = \lr{\prod \limits_{\tau=0}^{L_t-1} U_{\tau, \vec{x}, 0}}^{1/L_t}$ is the $L_t$'th matrix root of the Polyakov loop at spatial lattice site $\vec{x}$, that is, a matrix that satisfies $\bar{U}_{\vec{x},0}^{L_t} = \prod \limits_{\tau=0}^{L_t-1} U_{\tau, \vec{x}, 0}$. Starting with say $\omega_{0,\vec{x}} = I$ ($SU\lr{3}$ identity element), the equation (\ref{eq:sun_static_gauge}) allows to find all $\omega_{\tau,\vec{x}}$ for each $\vec{x}$.

After fixing the static gauge, we calculate the sum $S = \sum\limits_{\tau = 0}^{L_t - 1} U_{\tau,\vec{x},i}$ of $SU\lr{3}$ link matrices for each spatial lattice site $\vec{x}$ and spatial direction $i$, and solve the above minimization problem to find the $\tau$-averaged $SU\lr{3}$ link $\bar{U}_{\vec{x},i}$. Finally, we set $U_{\tau,\vec{x},i} = \bar{U}_{\vec{x},i}$ for all $\tau$.

We also tried a more gauge-invariant prescription where time-like links are parallel transported to a single time slice (e.g. $\tau = 0$) to calculate the averages. Such a prescription yields results that are very similar to the one obtained in the static gauge, but suffers from the ambiguity of choosing a particular time slice where the averages are calculated.

\subsection{The effect of static projection on meson correlators with different quantum numbers}
\label{subsec:su3:all_meson_correlators}

Before discussing the Hamiltonian-based calculations of real-time correlators and spectral functions in the static approximation, it is instructive to understand how well static approximation works for Euclidean-time meson correlators in lattice QCD. It is important that we don't need an implementation of the fermionic Hamiltonian to check the accuracy of the static approximation. Since static projection only affects gauge fields, we can measure meson correlators on static gauge field configurations using exactly the same code (with same parameters) that is used to measure the full meson correlators. In this work, we use the \texttt{Gen2L} ensembles by the \texttt{FASTSUM} collaboration \cite{FASTSUMGen2L,Glesaaen:2007.04188,Aarts:2209.14681} with temperatures $T = \lrc{380, 507, 760} \MeV$, and measure meson correlators using the \href{https://gitlab.com/fastsum/openqcd-hadspec}{\texttt{OpenQCD-hadspec}} code with clover-improved Wilson-Dirac operator on anisotropic lattices.

\begin{figure}[h!tpb]
    \includegraphics[width=0.5\textwidth]{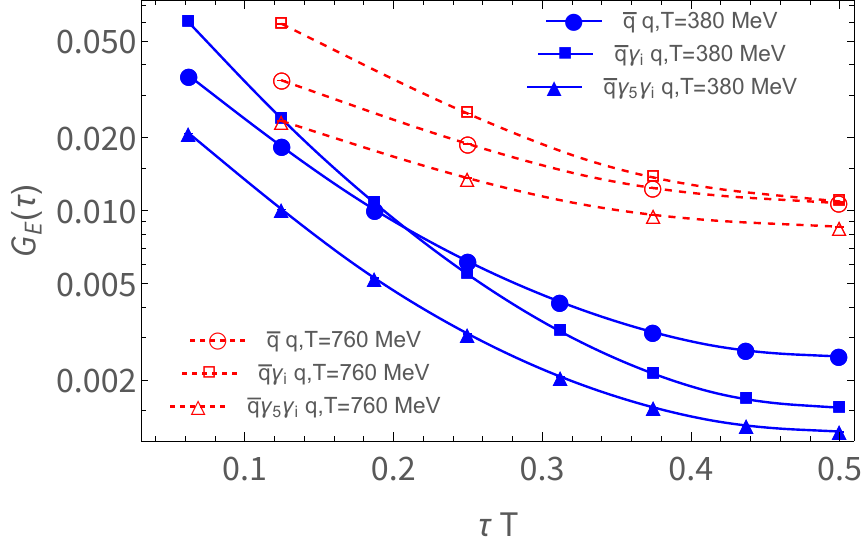}
    \caption{Raw data for the full quantum Euclidean-time meson correlators with different quantum numbers and for different temperatures. The points with $\tau=0$ are not shown because of contact term divergences.}
    \label{fig:GE_raw}
\end{figure}

On Fig.~\ref{fig:GE_raw}, we show the raw lattice results for several different meson correlators at different temperatures. We exclude the point with $\tau = 0$ to discard very large contact-term contributions that often change the sign of the correlator.

The ratios of Euclidean meson correlators calculated in the static approximation to the corresponding full quantum results are shown on Fig.~\ref{fig:GE_ratios_all}. For reference, we also show similar ratios between the free quark results and the full quantum result. 

All plots suggest that the ratios become closer to one at high temperatures, hence static approximation becomes more and more precise. However, a comparison with the free quark cases suggests that in some cases the deviation between the static approximation results and the full quantum results are larger than for free quarks. We conclude that at finite temperatures the success of the static approximation is observable-dependent. Importantly, static approximation is better than the free quark approximation for correlators of local vector and axial currents $\bar{q} \gamma_i q = \hat{q}^{\dag} \gamma_0 \gamma_i \hat{q}$ and $\bar{q} \gamma_5 \gamma_i q = \hat{q}^{\dag} \gamma_0 \gamma_5 \gamma_i \hat{q}$, as well as for the pion correlator with $O = \bar{q} \gamma_5 q$.

\begin{figure*}[h!tpb]
\includegraphics[width=0.45\textwidth]{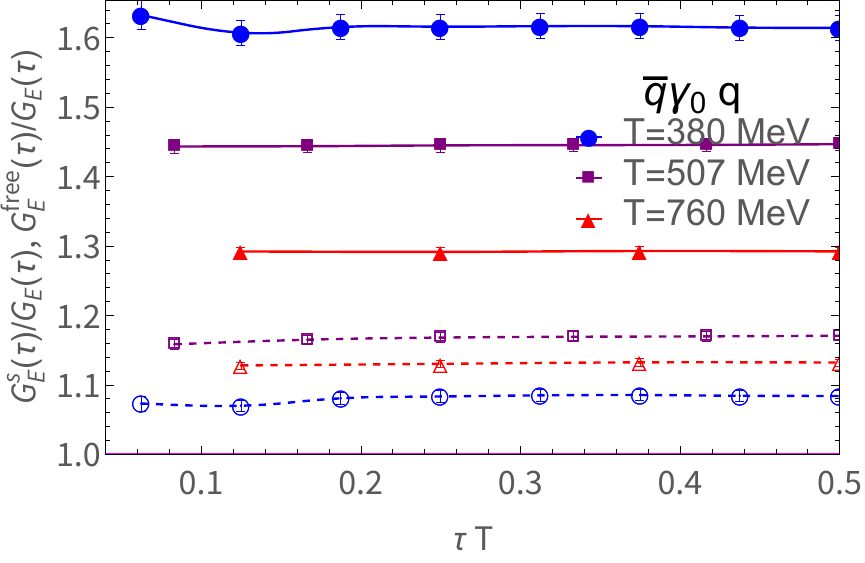}
\includegraphics[width=0.45\textwidth]{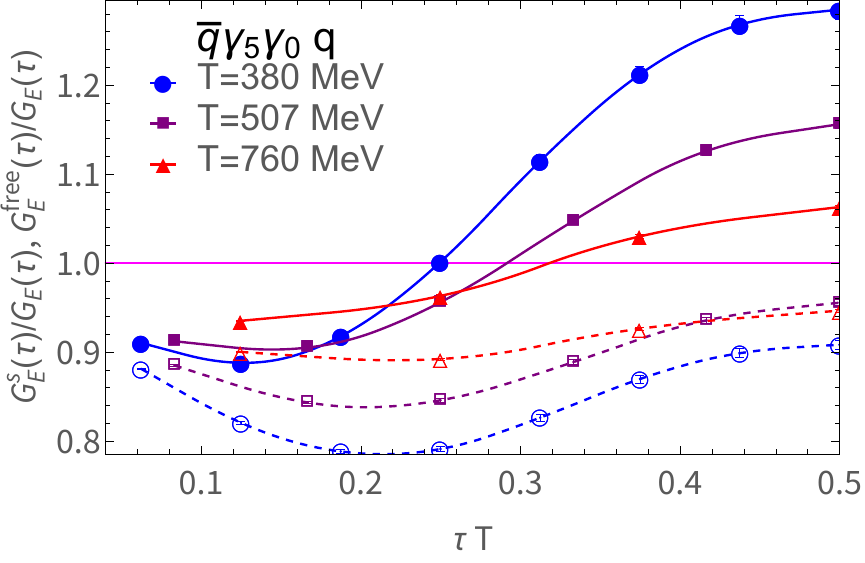}\\
\includegraphics[width=0.45\textwidth]{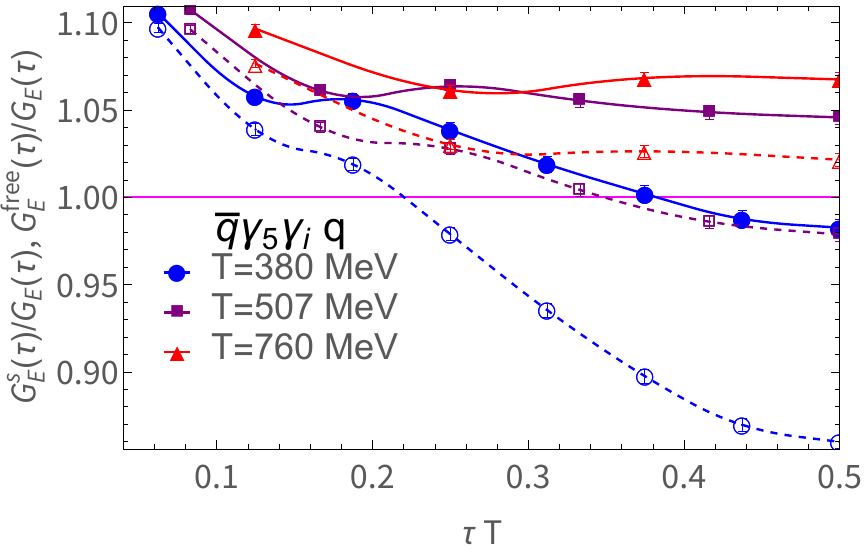}
\includegraphics[width=0.45\textwidth]{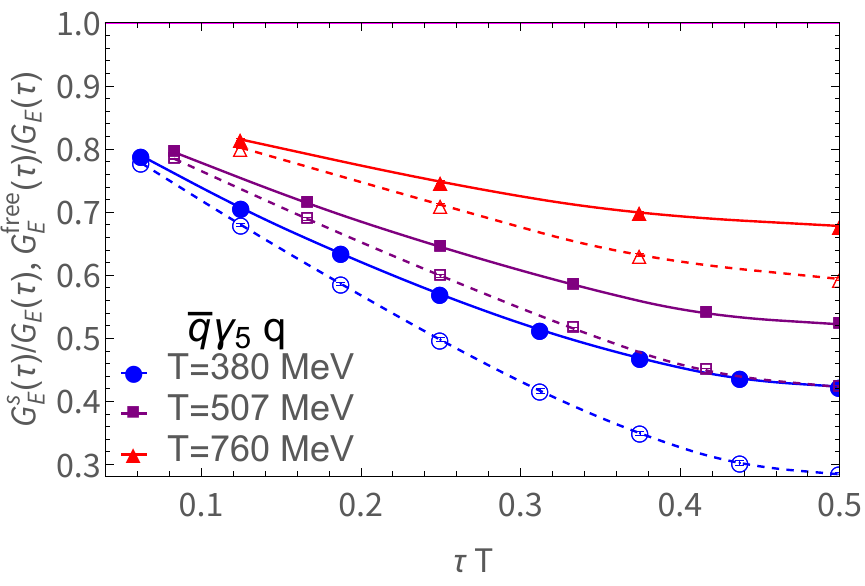}\\
\includegraphics[width=0.45\textwidth]{GE_ratios_qgiq.pdf}
\includegraphics[width=0.45\textwidth]{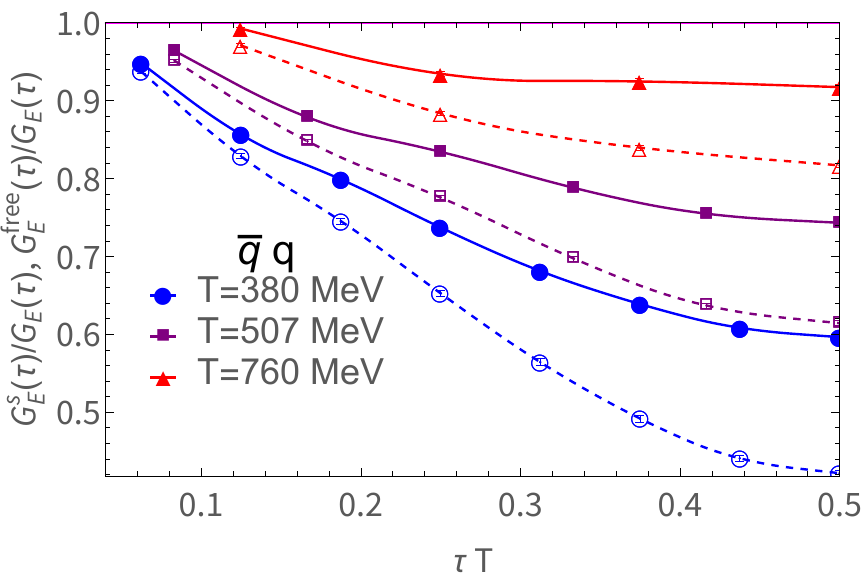}\\
\includegraphics[width=0.45\textwidth]{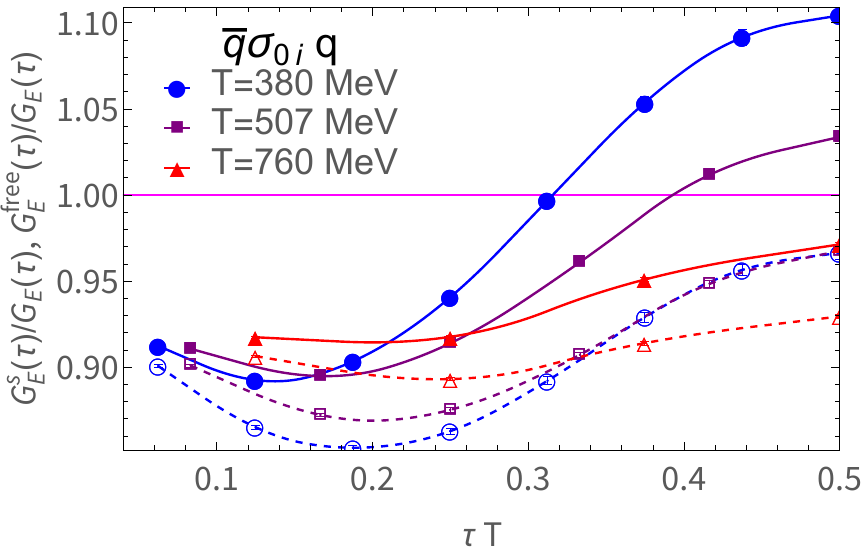}
\includegraphics[width=0.45\textwidth]{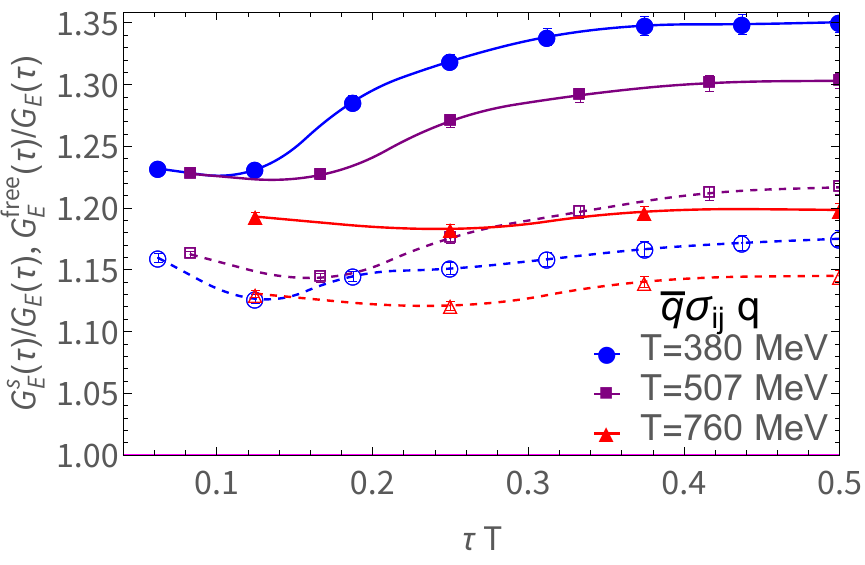}\\
\caption{Ratios of Euclidean-time meson correlators calculated in the static approximation (solid lines and  marker plots) and with free quarks (dashed lines and empty marker plots) to their full quantum counterparts, for different operators that span the full basis in the $4 \times 4$ spin matrix space.}
\label{fig:GE_ratios_all}
\end{figure*}

\subsection{Wilson-Dirac Hamiltonian}
\label{subsec:lqcd:hWD}

Here we describe the Wilson-Dirac single-particle Hamiltonian used in our work to calculate static approximation to spectral functions on \texttt{FASTSUM} \texttt{Gen2L} configurations. The Hamiltonian is directly derived from the clover-improved Wilson-Dirac operator on anisotropic lattices used by the \texttt{FASTSUM} collaboration \cite{Glesaaen:2007.04188,Aarts:2209.14681}:
\begin{eqnarray}
\label{eq:lqcd:hWD}
    h = \gamma_0 \lr{ m_0 + \frac{3}{u_t \, \gamma_f} + C_{\vec{x}} } \delta_{\vec{x},\vec{y}}  
    - \\ \nonumber - \frac{\gamma_0}{u_t \, \gamma_f}  \sum_{k=1}^{3} \lr{ \lr{ \frac{1 - \gamma_k}{2} } U_{\vec{x},k} \delta_{\vec{x}+\vec{e}_k, \vec{y}}  +  \lr{ \frac{1 + \gamma_k}{2} } U^{\dagger}_{\vec{y}, k} \delta_{\vec{x}-\vec{e}_k, \vec{y}} } ,
\end{eqnarray}
where
\begin{itemize}
\item $m_0$ is the bare mass,
\item $\{ \gamma_0, \gamma_k \}$ are Euclidean gamma matrices,
\item $\vec{e}_k$ is the basis vector on the lattice that corresponds to the shift by one lattice spacing in the direction $k$,
\item $U_{\vec{x}, k} \in SU(3)$ are spatial gauge link variables associated with links originating from a spatial lattice site $\vec{x}$,
\item $u_t$ and $u_s$ are the temporal and spatial tadpole improvement factors,
\item $\gamma_f$ is the bare fermionic anisotropy, defined in terms of the bare gluonic anisotropy $\xi_0$, and ratio of the gluonic and fermionic anisotropies $\nu$, which is related by $\gamma_f = \xi_0 / \nu$,
\end{itemize}
The clover term is given by
\begin{eqnarray}
\label{eq:lqcd:clover_term}
    C_{\vec{x}}  = -\frac{c_R}{2 \xi_0 \, u_t \, u_s^3} \sum_{1\le j < k \le 3} i \sigma_{j k} F_{\vec{x}, j k} , 
    \nonumber \\
    \sigma_{\mu \nu} = \frac{i}{2}[\gamma_{\mu}, \gamma_{\nu}],
\end{eqnarray}
where $C_R$ is the anisotropic spatial clover parameter. Temporal clover terms are considered a part of the discretization of the time derivative, and hence do not enter the Hamiltonian. The field strength tensor $F_{\vec{x}, \mu \nu}$ is calculated as
\begin{eqnarray}
\label{eq:lqcd:field_strength}
    F_{\vec{x},\mu \nu} = \frac{1}{8} \lr{Q_{\vec{x}, \mu \nu} - Q_{\vec{x}, \nu \mu} }, \\
    Q_{\vec{x},\mu \nu} 
    = 
    U_{\vec{x}, \mu} U_{\vec{x} + \vec{e}_{\mu}, \nu} U^{\dagger}_{\vec{x} + \vec{e}_{\nu}, \mu} U^{\dagger}_{\vec{x}, \nu} 
    + \nonumber \\ + 
    U_{\vec{x}, \nu} U^{\dagger}_{\vec{x} - \vec{e}_{\mu} + \vec{e}_{\nu}, \mu} U^{\dagger}_{\vec{x} - \vec{e}_{\mu}, \nu} U_{\vec{x} - \vec{e}_{\mu}, \mu} 
    - \\ \nonumber - 
    U^{\dagger}_{\vec{x} - \vec{e}_{\mu}, \mu} U^{\dagger}_{\vec{x} - \vec{e}_{\mu} - \vec{e}_{\nu}, \nu} U_{\vec{x} - \vec{e}_{\mu} - \vec{e}_{\nu}, \mu} U_{\vec{x} - \vec{e}_{\nu}, \nu} 
    +\\ \nonumber  + U^{\dagger}_{\vec{x} - \vec{e}_{\nu}, \nu} U_{\vec{x} - \vec{e}_{\nu}, \mu} U_{\vec{x} + \vec{e}_{\mu} - \vec{e}_{\nu}, \nu} U^{\dagger}_{\vec{x}, \mu} .
\end{eqnarray}
The values for the parameters used in the clover-improved Wilson-Dirac Hamiltonian for calculating real-time and Euclidean-time correlators are given in the table \ref{tab:parameters}.

\begin{table}[]
\centering
\begin{tabular}{ll ll}
\hline
Parameter & Value & Parameter & Value \\
\hline
$\nu$ & 1.265    & $\xi_0$ & 4.3     \\
$u_t$ & 1.0      & $u_s$   & 1.0 \\
$m_0$ & -0.084   & $c_R$   & 1.5893  \\
\hline
\end{tabular}
\caption{Values of the parameters of the clover-improved Wilson-Dirac Hamiltonian used in this work.}
\label{tab:parameters}
\end{table}

In order to match the Dirac operator definition used by the \texttt{FASTSUM} collaboration, the spatial links that enter the Wilson-Dirac Hamiltonian described above are also subject to $n=2$ steps of stout smearing with the spatial smearing parameter $\rho_s = 0.14$.

\subsection{Comparison of Euclidean-time meson correlators calculated in the Hamiltonian formulation and using the four-dimensional Wilson-Dirac operator.}
\label{subsec:lqcd:WD_vs_H}

In the main text, we derived the continuous-time expression for the Euclidean-time fermionic correlators in the background of static gauge fields that is based on the single-particle fermionic Hamiltonian:
\begin{eqnarray}
\label{eq:lqcd:GE_static_explicit}
 G^s_E\lr{\tau}
 = \vev{
    \tr\lr{
        O \, \frac{e^{-\tau h}}{1 + e^{-\beta h} z} \,
        O \, \frac{e^{-\lr{\beta - \tau} h} \, z}{1 + e^{-\beta h} z}
    }
    +
    \lr{\tr\lr{
        O \, \frac{1}{1 + e^{-\beta h} z}} \,
    }^2
    }
\end{eqnarray}
This expression is free of time discretization artefacts. In practice, we approximate the matrix functions $\frac{e^{-\tau h}}{1 + e^{-\beta h} z}$ and $\frac{e^{-\lr{\beta - \tau} h}}{1 + e^{-\beta h} z}$ using minmax polynomials with target precision of $10^{-5}$, and take the trace over the fermionic single-particle Hilbert space using $\mathbb{Z}_4$ stochastic estimators, with $O\lr{10}$ stochastic estimators per gauge field configuration being enough to reach the same level of statistical errors as for the averaging over gauge field configurations.

On the other hand, the same correlator can also be calculated using the full four-dimensional Wilson-Dirac operator in the same static gauge field background. Since Wilson-Dirac operator is based on finite difference approximations to time derivatives, a comparison between the continuous-time and the lattice correlators can be used to quantify time discretization artefacts. 

On Fig.~\ref{fig:GE_WD_vs_H} we plot the ratios of vector-vector correlators calculated using the four-dimensional Wilson-Dirac operator and continuous-time, Hamiltonian-based expressions on \texttt{FASTSUM} \texttt{Gen2L} configurations upon static projection. A comparison with the corresponding ratio plot for the $\bar{q} \gamma_i q$ operator in Fig.~\ref{fig:GE_ratios_all} demonstrates that time discretization artifacts for Wilson-Dirac operator at high temperatures (or small number of time slices $L_t$) can be even larger than the effect of removing the gauge fields altogether or projecting them to a static configuration.

Because of a numerically ill-defined nature of the spectral reconstruction procedure, even small systematic errors in Euclidean-time correlators can lead to large errors in the reconstructed spectral function. Furthermore, the Green-Kubo kernel $K\lr{\tau, w} = \frac{\cosh\lr{w \lr{\tau - \beta/2}}}{\sin\lr{w \beta/2}}$ assumes the continuous time $\tau$. It is therefore important to understand the implications of rather large time discretization artifacts for the spectral reconstruction procedure. We leave this investigation for further work.

\begin{figure}[h!tpb]
    \centering
\includegraphics[width=0.5\linewidth]{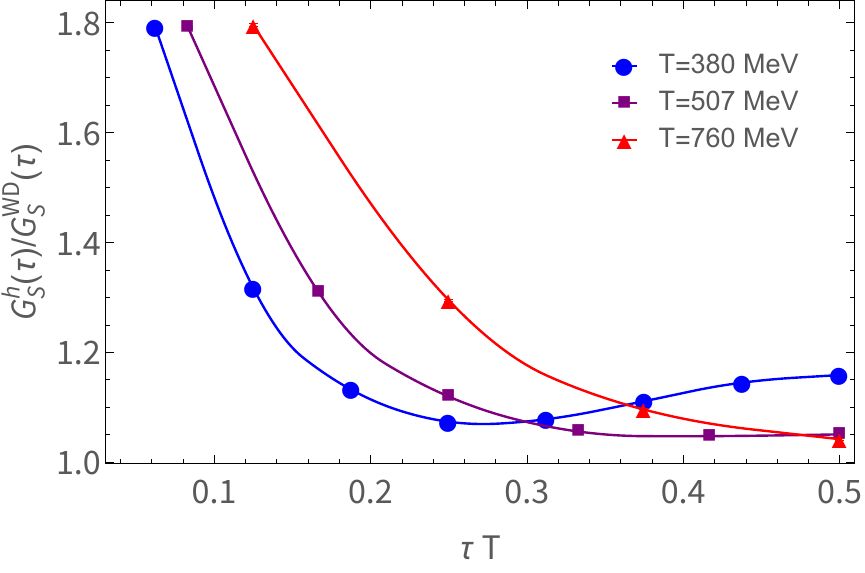}
    \caption{Ratios of static approximations to the vector-vector Euclidean correlator calculated using the four-dimensional Wilson-Dirac operator (as implemented in the \texttt{OpenQCD-fastsum} code) and the continuum-time, Hamiltonian-based expression in the main text with the Wilson-Dirac Hamiltonian (\ref{eq:lqcd:hWD}).}
    \label{fig:GE_WD_vs_H}
\end{figure}

\subsection{Calculation of real-time fermion correlators in lattice QCD}
\label{subsec:lqcd:GR_calculation}

As discussed in the main text, we calculate static approximations to spectral functions in lattice QCD from the Fourier transform of the real-time fermionic retarded correlator $G_R^s\lr{t}$ calculated in the background of static gauge fields:
\begin{eqnarray}
\label{eq:lqcd:GR}
	G_R\lr{t} = -i \vev{ \tr\lr{
        \lrs{n_z\lr{h} O n_z\lr{h}, O\lr{t}} e^{-\beta h} z} } .
\end{eqnarray}
For all temperatures $T = 760 \MeV$, $507 \MeV$, $380 \MeV$ that we consider in this work, the global center element $z$ in (\ref{eq:lqcd:GR}) is equal to $z = 1$, hence the Fermi distribution function $n\lr{h} \equiv n_{1}\lr{h} = \frac{1}{1 + e^{-\beta h}}$ of the single-particle Hamiltonian is a Hermitian matrix. In this case we can get rid of the commutator $\lrs{n_z\lr{h} O n_z\lr{h}, O\lr{t}}$ in the expression (\ref{eq:lqcd:GR}), which reduces the computational cost of calculating $G_R\lr{t}$ by a factor of two:
\begin{eqnarray}
\label{eq:lqcd:GR_z1}
	G_R\lr{t} = -i \, \im \vev{ \tr\lr{
        n\lr{h} \, O \, n\lr{h} \, e^{i h t} \, O \, e^{-i h t} \, e^{-\beta h}} } .
\end{eqnarray}
For practical lattice QCD calculations, the dimensionality of the fermionic single-particle Hamiltonian in (\ref{eq:lqcd:GR}) is still too large to allow for a direct calculation of (\ref{eq:lqcd:GR_z1}). Fortunately, it can still be calculated at relatively low computational cost. First of all, we use the approach similar to \cite{Borsanyi:0809.4711} and use stochastic estimators to calculate the trace in (\ref{eq:lqcd:GR_z1}). We use $\mathbb{Z}_4$ stochastic estimators, that is, vectors $\ket{\chi}$ with statistically independent, uniformly distributed components that take values in $\mathbb{Z}_4$ group: $\mathbb{Z}_4 = \lrc{1, +i, -1, -i}$. These vectors satisfy $\cev{ \ket{\chi}\bra{\chi} } = I$, where $\cev{\ldots}$ denotes averaging over random samples of $\chi$. We hence represent (\ref{eq:lqcd:GR_z1}) as
\begin{eqnarray}
\label{eq:lqcd:GR_z1_stochastic}
    G_R\lr{t} = -i \, \im \vev{ \cev{ \bra{\chi}
     O \, n\lr{h} \, e^{i h t} \, O \, e^{-i h t} \, e^{-\beta h} n\lr{h} \ket{\chi} } } . 
\end{eqnarray}
The quantity $\bra{\chi} n\lr{h} \, O \, n\lr{h} \, e^{i h t} \, O \, e^{-i h t} \, e^{-\beta h} \ket{\chi}$ under both expectation values can be calculated with $O\lr{L_s^3}$ complexity for each stochastic estimator $\ket{\chi}$. To this end, we first calculate 
\begin{eqnarray}
\label{eq:lqcd:GR_LR}
\ket{R} = e^{-\beta h} n\lr{h} \ket{\chi},
\quad
\ket{L} = n\lr{h} O \ket{\chi} .
\end{eqnarray}
The matrix functions $n\lr{h}$ and $e^{-\beta h} n\lr{h} = n\lr{-h}$ are approximated as minimax polynomials of $h$ with maximal deviation from the exact function less than $10^{-5}$. To set up these polynomials, we need to know the minimal $\epsilon_{min}$ and the maximal $\epsilon_{max}$ eigenvalues of $h$, which are calculated using Arnoldi algorithm implemented in the \href{https://github.com/m-reuter/arpackpp}{\texttt{arpackpp}} package. With a very good precision $\epsilon_{min} = -\epsilon_{max}$. We find that $\epsilon_{max} \approx 1.7 \, a_{\tau}^{-1}$ for the clover-improved Wilson-Dirac Hamiltonian (\ref{eq:lqcd:hWD}) with our range of temperatures, where $a_{\tau}^{-1} = 6.08 \, \mathrm{GeV}$ is the inverse lattice spacing in the time direction \cite{Aarts:2209.14681,Glesaaen:2007.04188}. The polynomial coefficients are obtained using the Remez algorithm. With our parameters and target precision, the polynomial orders range from $50$ for $T = 760 \MeV$ (where the Fermi distribution is a smooth function) to $90 \ldots 95$ for $T = 380 \MeV$, where the Fermi distribution changes faster around zero. To calculate $n\lr{h} \ket{\psi}$ given a polynomial approximation to $n\lr{h}$, we use Clenshaw recursive relations. With the Wilson-Dirac Hamiltonian $h$ in (\ref{eq:lqcd:hWD}) being a sparse matrix, the computational complexity of this step is proportional to the spatial volume times the polynomial degree. 

The vectors $\ket{R}$ and $\ket{L}$ are then evolved in real time to obtain $\ket{R\lr{t}} = e^{i h t} \ket{R}$ and $\ket{L\lr{t}} = e^{i h t} \ket{L}$. We move in time steps $\dt$ and use the relations $\ket{R\lr{t + \dt}} = e^{i h \dt} \ket{R\lr{t}}$ (and similarly for $\ket{L\lr{t}}$). The evolution factor $e^{i h \dt}$ is approximated in terms of its Taylor series expansion up to some finite order $N$:
\begin{eqnarray}
\label{eq:lqcd:time_step}
    \ket{R\lr{t + \dt}} = \sum\limits_{k=0}^N \frac{\lr{i \dt}^k}{k!} \, h^k \, \ket{R\lr{t}} .
\end{eqnarray}
This scheme is actually equivalent to $N$-th order Runge-Kutta integration for the equation $\partial_t \ket{R\lr{t}} = i h \, \ket{R\lr{t}}$. Again, for a sparse Hamiltonian matrix $h$ the application of a single time evolution factor requires the number of operation that is proportional to spatial volume times the approximation order $N$.

Finally, after calculating $\ket{R\lr{t}}$ and $\ket{L\lr{t}}$, we calculate the stochastic estimator contribution (\ref{eq:lqcd:GR_LR}) to real-time correlator (\ref{eq:lqcd:GR}) as
\begin{eqnarray}
\label{eq:lqcd:GR_final}
    G_R\lr{t} = -i \, \im \vev{ \cev{ \bra{L\lr{t}} \, O \,  \ket{R\lr{t}} } } . 
\end{eqnarray}

\subsection{Optimization of real-time evolution algorithm}
\label{subsec:lqcd:time_step_optimization}

After calculating the real-time correlator (\ref{eq:lqcd:GR}) over a time interval $t \in \lrs{0, t_{max}}$ with a time step $\dt$, we perform a Fourier transform to find the spectral function $S\lr{w}$. According to Nyquist theorem, we can reliably estimate $S\lr{w}$ up to frequencies $w_{max} = \frac{\pi}{\dt}$, and achieve the frequency resolution of $\Delta w = \frac{2 \pi}{t_{max}}$. Explicit expressions for $S\lr{w}$ in terms of the eigenspectrum of a single-particle Hamiltonian make it clear that $S\lr{w}$ vanishes for $w > 2 \epsilon_{max}$. For the static approximation to $S\lr{w}$ to serve as a model function in the MEM method, it is important to probe the entire frequency range where $S\lr{w}$ is not vanishing. Therefore, the time step $\dt$ should obey 
\begin{eqnarray}
 \dt \leq \frac{\pi}{2 \, E_{max}} ,
\end{eqnarray}
which yields $\dt \leq 0.93 \, a_t^{-1}$ for $E_{max} \approx 1.7 \, a_t^{-1}$. 

To take the full advantage of the static approximation to $S\lr{w}$, the frequency resolution for $S\lr{w}$ should be considerably smaller than the frequency resolution $\Delta w \sim \pi \, T$ of generic spectral reconstruction methods. It does not make sense to make $\Delta w$ too small, as static approximation eventually becomes invalid for very small frequency scales. With $\Delta w \lesssim 0.1\, T$, we estimate $t_{max} \gtrsim \frac{2 \pi \, L_t}{0.1} \, a_t^{-1} \approx 10^3$ for our lowest temperature value $T = 380 \, \MeV$ ($L_t = 16$). 

The worst-case error accumulated over the full time interval $t_{max}$ may be approximated by
\begin{align}
    \abs{ \lr{ \sum_{n=0}^{N} \frac{(i \epsilon_{max} \Delta t)^n}{n!} }^{\frac{t_{max}}{\Delta t}} - e^{i \epsilon_{max} t_{max}} } = \abs{ R_N } .
\end{align}
We want to choose $\Delta t$, $t_{max}$, and $N$ such that the remainder is smaller than some target precision (in this work we require $\abs{R_N} \leq 10^{-5}$), the inequalities on $t_{max}$ and $\dt$ are satisfied as discussed above, and the computational cost is minimal. The computational cost is dominated by the operation $h \ket{\psi}$ of multiplying an arbitrary vector $\ket{\psi}$ in fermionic single-particle Hilbert space by the sparse matrix of the single-particle fermionic Hamiltonian $h$. The number of these operations required to reach the maximal evolution time $t_{max}$ is
\begin{eqnarray}
\label{eq:lqcd:num_mult}
    N_{mult} = \frac{t_{max}}{\dt} \, N .
\end{eqnarray}
To achieve the target precision for $G_R\lr{t}$ in (\ref{eq:lqcd:GR_final}), one can either reduce the time step $\dt$ or increase the expansion order $N$. In general, increasing $N$ is more efficient. By comparing computational costs at different $\dt$ and $N$, we found that $\Delta t = 0.8$ and $N = 12$ is an optimal choice with $\abs{R_N} = 1.07 \times 10^{-5}$.

We note that evolution of only fermionic degrees of freedom is technically considerably simpler than the simultaneous evolution of fermions and gauge fields that is performed in classical-statistical approximation. In particular, the linearity of the fermionic evolution equations allows to implement higher-order Runge-Kutta methods very easily, as they become equivalent to higher-order Taylor series expansions of matrix exponentials.

\subsection{Quantum typicality of real-time evolution with Wilson-Dirac Hamiltonian in lattice QCD at finite temperature}
\label{subsec:lqcd:quantum_typicality}

The success of our numerical strategy to calculate the real-time correlator (\ref{eq:lqcd:GR}) using stochastic estimator representation (\ref{eq:lqcd:GR_z1_stochastic}) crucially depends on typical numbers of gauge configurations and stochastic estimators needed to calculate the expectation value in (\ref{eq:lqcd:GR_z1_stochastic}) with sufficiently small statistical errors. It is therefore important to understand the variation in the contributions of different stochastic estimators to (\ref{eq:lqcd:GR_z1_stochastic}).

In the left panel of Fig.~\ref{fig:SPF_stochastic} we show spectral functions obtained as Fourier transforms of individual contributions $2 \im \bra{\chi} O \, n\lr{h} \, e^{i h t} \, O \, e^{-i h t} \, e^{-\beta h} n\lr{h} \ket{\chi}$ to the expectation value (\ref{eq:lqcd:GR_final}), where the operator $O = \gamma_0 \, \gamma_i$ corresponds to the local (non-conserved) vector current operator. These plots suggest that the variation of the results for different stochastic estimators is comparable to that for statistically independent gauge configurations. Overall the variation is reasonably small in both cases, and is expectably larger towards low frequencies. We found that around $10$ stochastic estimators per configuration is enough to obtain static approximations to spectral functions with very small errors (see the plots in the main text). The fact that real-time evolution is very similar for different state vectors in the fermionic single-particle Hilbert space hints at the quantum typicality of fermion dynamics in QCD \cite{Bartsch:0902.0927}.

Let us note that if we apply the same strategy to free quarks, we get a very non-smooth spectral function which looks nothing like Fig.~\ref{fig:SPF_stochastic}, but rather as a collection of delta-functions. Thus random gauge field background is essential for the spectral function to be a continuous function that can be used as a model function in the MaxEnt method. 

\begin{figure*}[h!tpb]
\includegraphics[width=0.48\textwidth]{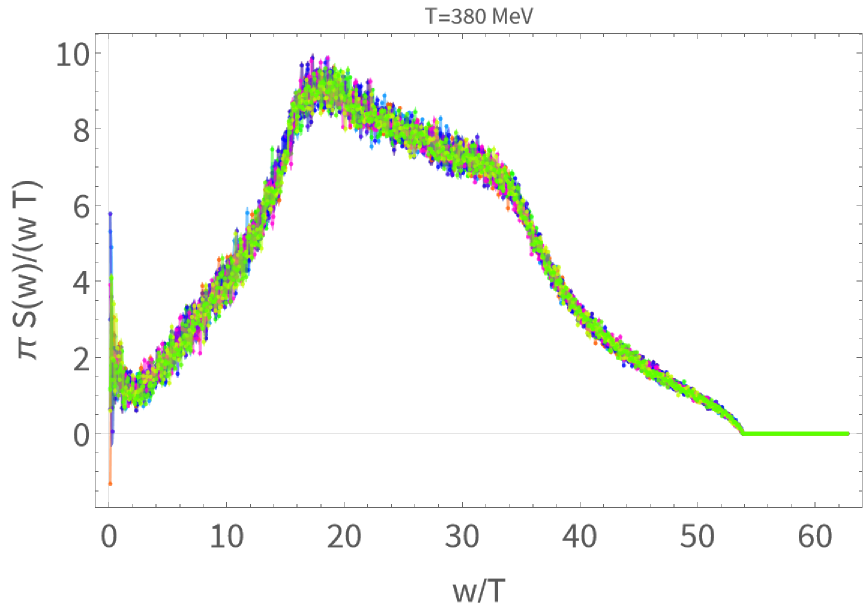}
\includegraphics[width=0.48\textwidth]{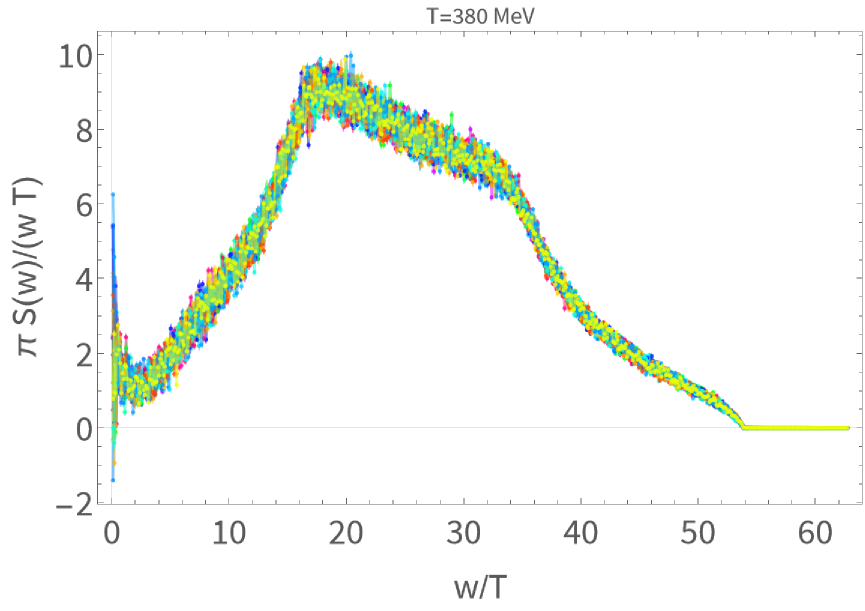}\\
\includegraphics[width=0.48\textwidth]{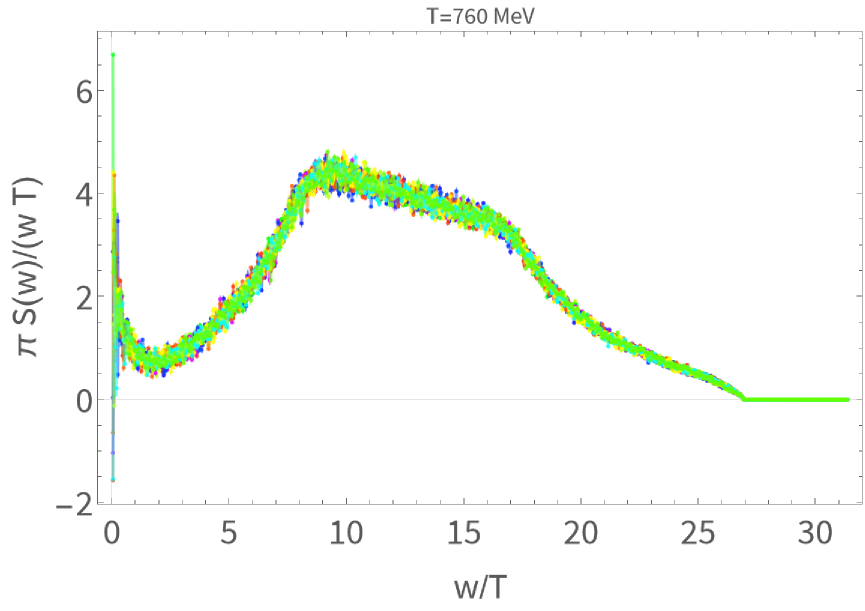}
\includegraphics[width=0.48\textwidth]{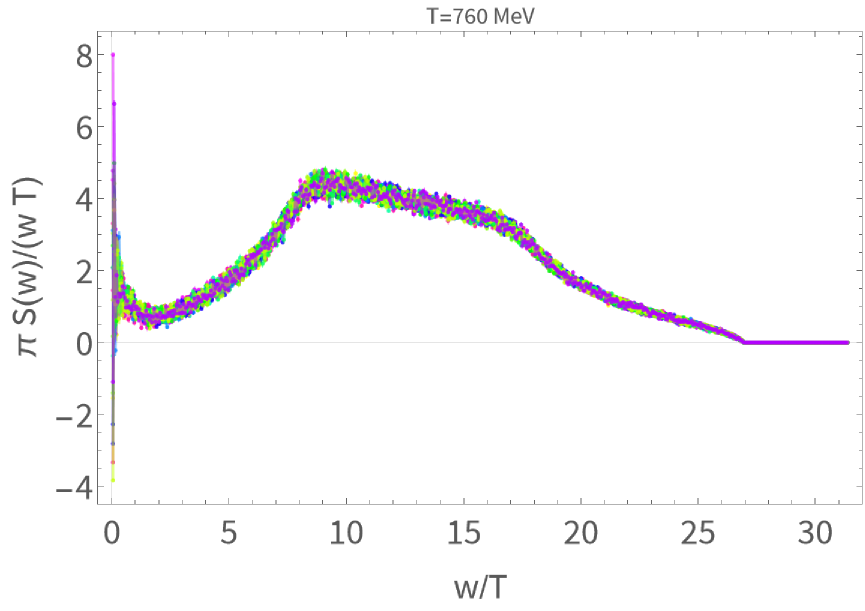}\\
\caption{Spectral functions obtained as Fourier transforms of individual contributions of statistically independent gauge configurations and stochastic estimators to the real-time correlator (\ref{eq:lqcd:GR_final}). On the left: for $10$ statistically independent gauge field configurations with $1$ stochastic estimator per configuration. On the right: for $10$ statistically independent stochastic estimators on a single randomly selected gauge field configuration.}
\label{fig:SPF_stochastic}
\end{figure*}

Finally, we note that our estimates of the zero-frequency limit of the spectral function yield the value of electric conductivity $\sigma\lr{w} = \frac{\pi \, S\lr{w}}{w}$ that is larger than the commonly quoted values $\sigma\lr{w}/T \approx 0.3 \ldots 0.6$ \cite{Nikolaev:2008.12326} by a factor of two or three. This difference is completely independent of the spectral reconstruction method, and can be easily traced to the full Euclidean vector-vector correlators produced by the \texttt{OpenQCD-hadspec} code on \texttt{FASTSUM} \texttt{Gen2L} configurations. It can be partly explained by the artifacts of using the non-conserved (local) vector current operators. \texttt{FASTSUM} ensembles have rather coarse spatial lattice spacing, that may lead to significant deviations of the vector current renormalization constant from unity (for example, $Z_V \approx 0.7$ for the coarse-lattice simulations of \cite{Aoki:hep-lat/0309185}). Over-smearing of the transport peak by spectral reconstruction methods such as Backus-Gilbert or MaxEnt may also contribute to this mismatch.

	
\end{widetext}


\end{document}